\documentclass[11pt]{article}
\pdfoutput=1
\usepackage{fullpage}
\usepackage{graphicx}
\usepackage{color}
\def\lsim{\mathrel{\rlap{\lower4pt\hbox{\hskip1pt$\sim$}}
    \raise1pt\hbox{$<$}}}         
\def\gsim{\mathrel{\rlap{\lower4pt\hbox{\hskip1pt$\sim$}}
    \raise1pt\hbox{$>$}}}         
\begin{document}
\begin{flushright}
UCB-NPAT-12-014,~
NT-LBL-12-017
\end{flushright}
\begin{center}
{\bf NEUTRINO ASTROPHYSICS} \\
 W. C. Haxton \\
 Department of Physics, University of California, Berkeley, \\
 and Lawrence Berkeley National Laboratory, \\
 MC-7300, Berkeley, CA 94720 \\
 email: haxton@berkeley.edu
\end{center}

\section{Introduction}
\noindent
The neutrino \cite{hh00} is an elementary particle that scatters only through the weak interaction, and consequently rarely interacts in matter.  Neutrinos are neutral, carry spin-1/2, and are members of the family of elementary particles called leptons.  Thus they differ from the quarks of the Standard Model (spin-1/2 particles which participate in strong and electromagnetic interactions) and from the other leptons, which are charged and thus interact electromagnetically.   Neutrinos 
and their antiparticles come in three types $-$ or flavors $-$ labeled according to the charged partners, the electron, muon, or tauon, that accompany neutrino production in charge-changing weak interactions.  The most familiar such reaction is $\beta$ decay
\[ (N,Z) \rightarrow (N-1,Z+1) +  e^- + \bar{\nu}_e \]
in which a nucleus containing N neutrons and Z protons decays to a lighter nucleus by converting
a neutron to a proton, with the emission of an electron and an electron antineutrino.  Indeed, it was the apparent absence of energy conservation in nuclear $\beta$ decay that first lead Wolfgang Pauli, 
more than 80 years ago, to suggest that some undetected neutral particle (the $\nu_e$) must be escaping from nuclear $\beta$ decay experiments.

Neutrinos play a very special role in astrophysics \cite{bahcallbook}.  First, they are the direct byproducts of the nuclear reaction chains by which stars generate energy: each solar conversion of four protons into helium produces two neutrinos, for a total of $\sim$ 2 $\times$ 10$^{38}$ neutrinos each second.  The resulting flux is observable on Earth.  These neutrinos carry information about conditions deep in the solar core, as they typically leave the Sun without further interacting.   They also provide experimentalists with opportunities for testing the properties of neutrinos over long distances.  Second, they are produced in nature's most violent explosions, including the Big Bang, core-collapse supernovae, and the accretion disks encircling supermassive black holes.  Recent discoveries of neutrino mass show that primordial neutrinos comprise a small portion of the ``dark matter" that influences how large-scale structure -- the pattern of voids and galaxies mapped by astronomers -- formed over cosmological times.  Third, they dominate the cooling of many astrophysical objects, including young neutron stars and the degenerate helium cores of red giants.  Neutrinos can be radiated from deep within such bodies, in contrast to photons, which are trapped within stars, diffusing outward only slowly.  Finally, neutrinos are produced in our atmosphere and elsewhere as secondary byproducts of cosmic-ray collisions.  Detection of these neutrinos can help constrain properties of the primary cosmic ray spectrum.  Neutrinos produce
by reactions of ultra-high-energy cosmics rays can provide information on otherwise inaccessible
cosmic accelerators.

Neutrinos also mediate important astrophysical processes.  While the site of the r-process -- the
rapid-neutron-capture process thought to be responsible for the nucleosynthesis of
about half of the neutron-rich nuclear species heavier than iron -- remains uncertain, 
the needed neutrons may be generated by the extraordinary neutrino fluxes found in core-collapse
supernovae.  Possible r-process sites include the neutrino-rich ``neutrino winds" that
blow off the proto-neutron star surface as well as the ${}^4$He zones of metal-poor
supernovae, where 
neutrons are produced by neutrino-induced spallation reactions.
Supernova neutrinos can also directly synthesize certain rare nuclei, 
such as ${}^{11}$B and ${}^{19}$F.

Nuclear and particle physicists are exploiting astrophysical neutrino fluxes to do important tests of the Standard Model.  These tests include neutrino oscillations (the process by which a massive neutrino can be produced in one flavor state but detected later as a neutrino with a different flavor), neutrino decay, 
the cosmological effects of neutrino mass, and searches for nonzero neutrino electromagnetic moments.

\section{Solar Neutrinos}
\noindent
The first successful effort to detect neutrinos from the Sun began four decades ago.  Ray Davis, Jr. and his collaborators constructed a 650-ton detector in the Homestake Gold Mine, one mile beneath
Lead, South Dakota \cite{davis}.  This radiochemical detector, based on the chlorine-bearing cleaning fluid C$_2$Cl$_4$, was designed to capture about one of the approximately 10$^{18}$  high-energy neutrinos that penetrated it each day -- the rest
passed through the detector, without interacting.  The neutrino-capture reaction was 
inverse-electron-capture
\[ {}^{37}\mathrm{Cl} + \nu_e^{solar} \rightarrow {}^{37}\mathrm{Ar} + e^-. \]
The product of this reaction, ${}^{37}$Ar, is a noble-gas isotope with a half life of about one month.  It can be efficiently removed from a large volume of organic fluid by a helium gas purge, then counted in miniature gas proportional counters as  ${}^{37}$Ar decays back to ${}^{37}$Cl.  Davis typically exposed his detector for
about two months, building up to nearly the saturation level of a few dozen argon
atoms, then purged the detector to determine the number of solar neutrinos captured during this period.

Within a few years it became apparent that the number of neutrinos detected was only about one-third that
predicted by the standard solar model (SSM) \cite{bahcallbook,SHP}, that is, the model of the Sun based on the standard theory of main sequence stellar evolution. Some initially attributed this ``solar neutrino problem" to uncertainties in the SSM: As the flux of neutrinos most important to the Davis detector vary as $\sim$ $T^{22}_c$, where $T_c$ is the solar core temperature, a 5\% theory uncertainty in $T_c$ could explain the discrepancy.  In fact, the correct explanation for the discrepancy proved much more profound.  Davis was awarded the 2002 Nobel Prize in Physics for the chlorine experiment.

\begin{figure}
\begin{center}
\includegraphics[width=12cm]{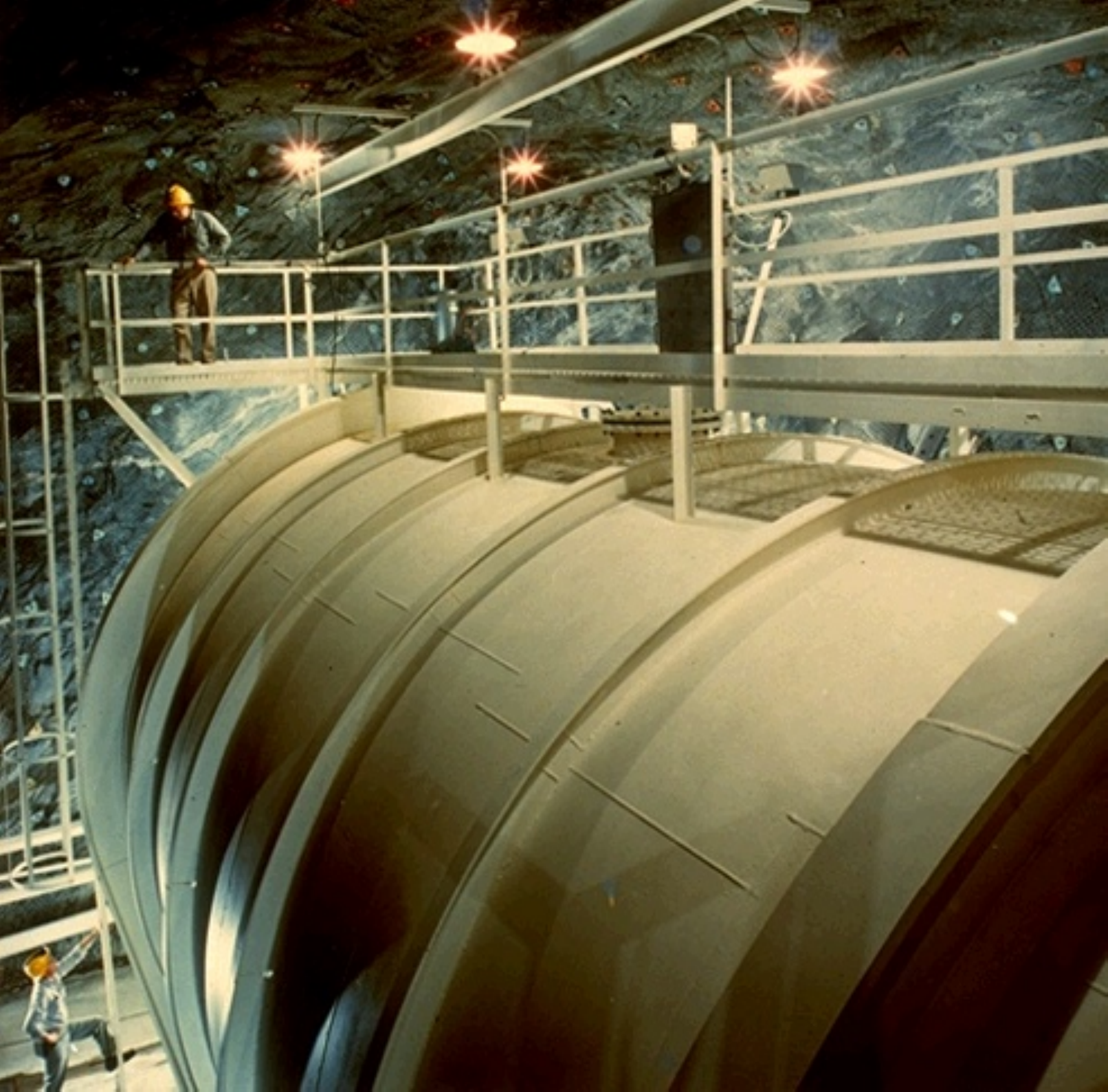}
\end{center}
\caption{ The Homestake Mine's chlorine detector, which Ray Davis Jr. and colleagues
operated for over three decades.}
\label{fig:one}
\end{figure}

The solar neutrino problem stimulated a series of follow-up experiments to measure the different
components of the solar neutrino flux and to determine the source of the Cl experiment
discrepancy.  The SAGE and GALLEX/GNO  experiments, radiochemical detectors similar to Cl,  but using ${}^{71}$Ga as a target, were designed to measure the flux of neutrinos from
the dominant low-energy branch of solar neutrinos, the pp neutrinos.  The first detector to measure neutrinos event by event, recording neutrino interactions in real time, was the converted proton decay detector Kamiokande.   The detector contained three kilotons of very pure water, with solar neutrinos scattering off the electrons within the water.  Phototubes surrounding the tank recorded the ring of Cerenkov radiation produced by the recoiling relativistic electrons.  Kamiokande measured the high-energy neutrinos most important to the Davis detector, and thus confirmed the deficit that Davis had originally observed.  New and very massive water (Super-Kamiokande) and heavy-water (Sudbury Neutrino Observatory (SNO)) detectors were constructed.  Finally, Borexino, a detector using liquid scintillator, was constructed to measure 
low-energy solar neutrino branches in real time.  These experiments -- most particularly SNO, 
because of its multiple detection channels sensitive to different combinations of neutrino types -- showed that solar neutrinos were not missing, but rather hidden by a change of flavor occurring during their transit from the Sun to the Earth, as will be described later in this chapter.   

Of these experiments, SAGE, Super-Kamiokande, and Borexino remain in operation.

\subsection{The Standard Solar Model}
\begin{figure}
\begin{center}
\includegraphics[width=17cm]{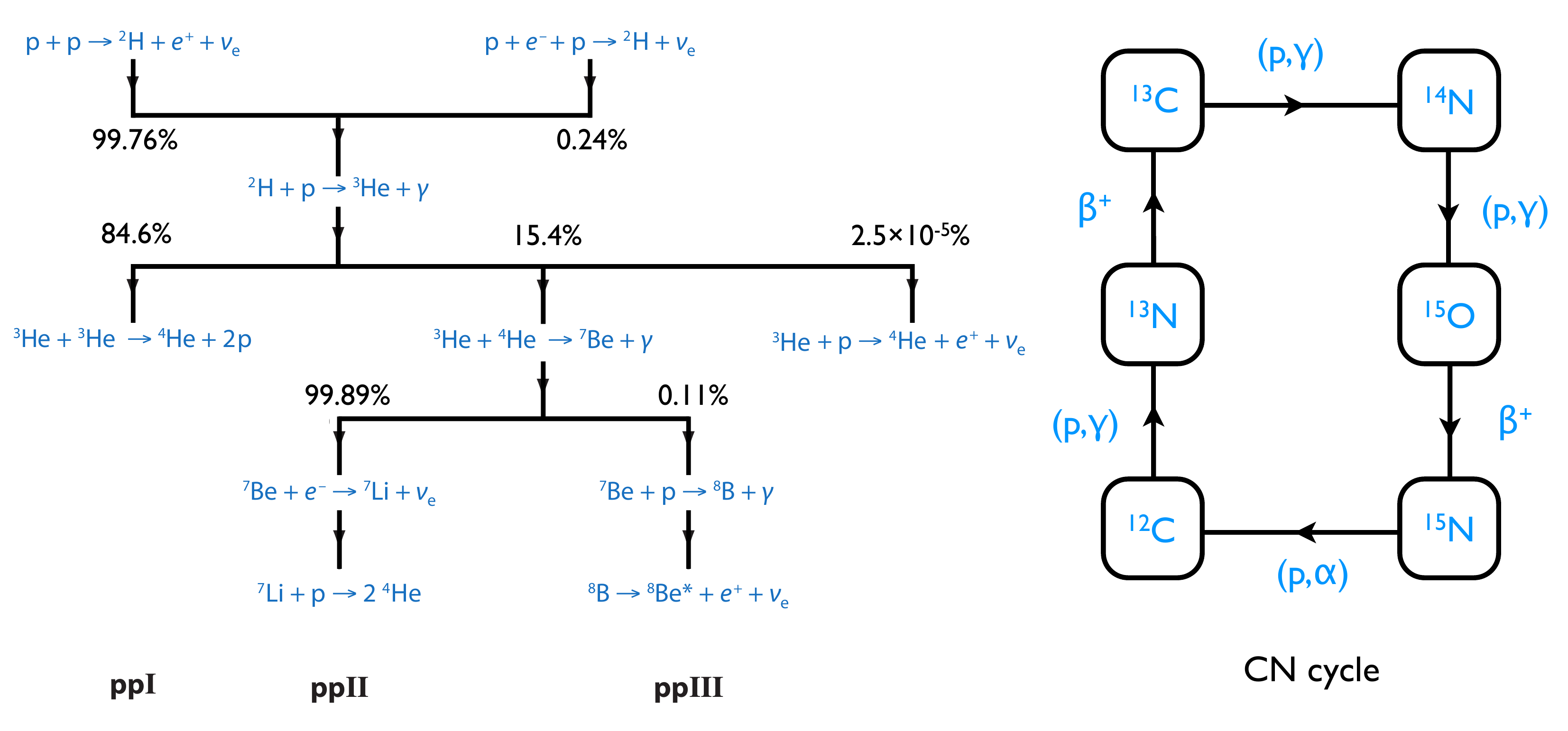}
\end{center}
\caption{ Left panel: The three principle cycles comprising the pp chain (denoted ppI, ppII, and ppIII).
Associated neutrinos ``tag" the three branches.  The SSM branching ratios come from the
GS98-SFII SSM \cite{SHP}.  Also shown is the minor branch $^3$He+p$\rightarrow {}^4$He+e$^+$+$\nu_e$,
which generates the Sun's most energetic neutrinos.  Right panel: The CN I cycle, which produces 
solar neutrinos from the $\beta$ decays of ${}^{13}$N and ${}^{15}$O.}
\label{fig:ppCNO}
\end{figure}

The Sun belongs to a class of ``main sequence'' stars that derive their energy from burning protons to He in their cores.  The SSM employs the standard theory of main-sequence stellar evolution, calibrated by the many detailed measurements only possible for the Sun, to follow the Sun from the onset of thermonuclear reactions 4.6 Gyr ago to today, thereby determining the present-day temperature and composition profiles of the solar core.  These profiles govern solar neutrino production and other properties of the modern Sun.  The SSM is based on four basic assumptions:
    
\begin{itemize}
\item The Sun evolves in hydrostatic equilibrium, maintaining a local balance between the gravitational force and the pressure gradient.  To describe this condition in detail, one must  specify the electron-gas equation of state as a function of temperature, density, and composition.  This requires attention to such issues
as the incomplete ionization of metals, the contribution of radiation to pressure, and the influence of
screening.
\item Energy is transported by radiation and convection.  While the solar envelope is convective, radiative transport dominates in the core region where thermonuclear reactions take place.  In addition to
elementary processes such as the scattering of photons off electrons and fully ionized H and He, more complex
processes such as bound-free scattering off metals are important contributors to the opacity in the
Sun's interior regions.
\item The Sun produces its energy by fusing protons into ${}^4$He,

\[ 2e^-+4\mathrm{p} \rightarrow {}^4\mathrm{He}  + 2 \nu_e +26.73 \mathrm{~MeV} \]
via the pp chain (99\%) and CN I cycle reactions of Fig. \ref{fig:ppCNO}.
The Sun is
a large but slow reactor: the core temperature, $T_c \sim  1.5 \times 10^7$ K, results in typical center-of-mass energies for reacting particles of $\sim$ 10 keV, much less than the Coulomb barriers inhibiting
charged-particle nuclear reactions.  Solar reaction rates are so slow that laboratory measurements
of these rates are not, in most cases, feasible at solar energies, but instead must be made at higher
energies and then extrapolated to the solar Gamow peak. 
\item The model is constrained to produce today's solar radius, mass, and luminosity.  An important assumption of the SSM is that the proto-Sun passed through a highly convective phase, rendering the Sun uniform in composition
until main-sequence burning began.  The initial composition by mass is conventionally divided into hydrogen (X$_\mathrm{ini}$),
helium (Y$_\mathrm{ini}$), and everything else (the metals, denoted Z$_\mathrm{ini}$),
with X$_\mathrm{ini}$+Y$_\mathrm{ini}$+Z$_\mathrm{ini}$=1.  The relative abundances of the metals
are determined from a combination of meteoritic and solar photospheric data.  The absolute 
abundance Z$_\mathrm{ini}$ can be taken from the modern Sun's surface abundance Z$_\mathrm{S}$,
after corrections for the effects of diffusion over 4.6 Gyr of solar evolution. 
Finally, Y$_\mathrm{ini}$/X$_\mathrm{ini}$ is adjusted along with $\alpha_\mathrm{MLT}$,
a parameter describing solar mixing, until the model reproduces the modern Sun's
luminosity and radius.  The resulting ${}^4$He/H mass fraction ratio is typically 0.27 $\pm$ 0.01, which can be compared to the Big-Bang value of 0.23 $\pm$ 0.01, showing that the
Sun was formed from previously processed material. 
\end{itemize}
  
Three cycles with quite different temperature dependences, reflecting the relative ease or difficulty of Coulomb barrier penetration, comprise the pp chain of Fig.~\ref{fig:ppCNO}.  The competition between the cycles is very sensitive to the solar core temperature $T_c$.  The initial interest in solar neutrinos came from the observation that each of the three cycles is associated with a characteristic neutrino.  Thus, by measuring solar neutrinos, specifically the pp, ${}^7$Be, and ${}^8$B neutrinos, one can determine the relative importance of the ppI, ppII, and ppIII cycles, and consequently determine $T_c$ to an accuracy of $\lsim$ 1\%.  

The neutrino-producing reactions of the pp chain and CN I cycle are summarized in Table \ref{tab:one}.  
The $\beta$ decay sources produce neutrinos with a continuous spectra, while the electron
capture reactions produce lines with widths $\sim$ 2 keV characteristic of the temperature of the solar core.
The Table shows two solar models, denoted GS98-SFII and AGSS09-SFII, which differ in their
assumptions about solar surface metallicity due to the use of 1D or 3D models, respectively,
to interpret photospheric absorption lines.  The predictions of the  higher metallicity ($\sim$ +30\%) 
GS98-SFII SSM are generally in excellent agreement with solar helioseismic properties, including
interior sound speeds and the location of the base of the convective zone.  This is not the
case for the AGSS09-SFII SSM, which nevertheless uses a more sophisticated treatment of
the photosphere.   The unresolved conflict between SSMs that agree with our best description
of the solar interior and those that employ our best model of the solar surface is known as the
solar abundance problem. 

The last line of the Table shows that the neutrino flux predictions of the two SSMs are almost
identical in the quality of their agreement with fluxes derived from experiment.  A best-fit would
be obtained for a metallicity Z$_\mathrm{ini}$ intermediate between GS98 and AGSS09.

\begin{table}
\caption{SSM neutrino fluxes from the GS98-SFII (high Z) and AGSS09-SFII (low Z) SSMs, which
differ in their assumptions about photospheric metallicity.  In cases where associated uncertainties
are asymmetric, an average has been used.  The solar values
come from a luminosity-constrained analysis of all available data by the Borexino Collaboration.
For references see \cite{HRS}.}
\vspace{0.4cm}
\label{tab:one}
\begin{tabular}{lccccc}
\hline \hline
 $\nu$ flux & E$_\nu^\mathrm{max}$ (MeV) & GS98-SFII & AGSS09-SFII & Solar & units \\
\hline
p+p$\rightarrow^2$H+e$^+$+$\nu$ & 0.42 & $5.98(1 \pm 0.006)$ & $6.03(1 \pm 0.006)$ & $6.05(1^{+0.003}_{-0.011})$ & 10$^{10}$/cm$^2$s \\
p+e$^-$+p$\rightarrow^2$H+$\nu$ & 1.44 & $1.44(1 \pm 0.012)$ & $1.47(1 \pm 0.012)$ & $1.46(1^{+0.010}_{-0.014})$ & 10$^8$/cm$^2$s\\
$^7$Be+e$^-$$\rightarrow^7$Li+$\nu$ & 0.86 (90\%) & $5.00(1 \pm 0.07)$ & $4.56(1 \pm 0.07)$ & $4.82(1^{+0.05}_{-0.04})$ & 10$^9$/cm$^2$s\\
 & 0.38 (10\%) & & & & \\
$^8$B$\rightarrow^8$Be+e$^+$+$\nu$ & $\sim$ 15 & $5.58(1 \pm 0.14)$ & $4.59(1 \pm 0.14)$ & $5.00(1\pm 0.03)$ & 10$^6$/cm$^2$s\\ 
${}^3$He+p$\rightarrow^4$He+e$^+$+$\nu$  & 18.77 & $8.04(1 \pm 0.30)$ & $8.31(1 \pm 0.30)$ & --- & 10$^3$/cm$^2$s\\
$^{13}$N$\rightarrow^{13}$C+e$^+$+$\nu$  & 1.20 & $2.96(1 \pm 0.14)$ & $2.17(1 \pm 0.14)$ &$\leq 6.7$ & 10$^8$/cm$^2$s\\ 
$^{15}$O$\rightarrow^{15}$N+e$^+$+$\nu$  & 1.73 & $2.23(1 \pm 0.15)$ & $1.56(1 \pm 0.15) $ &$\leq 3.2$ & 10$^8$/cm$^2$s\\ 
$\chi^2/P^\mathrm{agr}$ & & 3.5/90\% & 3.4/90\% & & \\
\hline \hline
\end{tabular}
\end{table}

\subsection{SNO , Super-Kamiokande, and Borexino}
Solar neutrino detection requires the combination of a large detector
volume (to provide the necessary rate of events), very low backgrounds (so
that neutrino events can be distinguished from backgrounds due to cosmic
rays and natural radioactivity), and a distinctive signal.  The first requirement
favors detectors constructed from inexpensive materials and/or materials having
large cross sections for neutrino capture.  The second generally requires
a deep-underground location for the detector, with sufficient rock overburden to
attenuate the flux of penetrating muons
produced by cosmic ray interactions in the atmosphere.  It also requires very careful attention to detector cleanliness,
including tight limits on dust or other contaminants that might introduce
radioactivity, use of low-background construction materials, control of radon, and
often the use of fiducial volume cuts so that the outer portions of a detector
become a shield against activities produced in the surrounding rock walls.

There are several possible detection modes for solar neutrinos,
interesting because of their different sensitivities to flavor.
The early radiochemical experiments using ${}^{37}$Cl and ${}^{71}$Ga
targets were based  on the charged current weak reaction 
\[ \nu_e + (N,Z) \rightarrow e^- + (N-1,Z+1) \]
where the signal for neutrino absorption is the growth over time of 
very small concentrations of the
daughter nucleus $(N-1,Z+1)$ in the detector.  As the spectrum of solar
neutrinos extends only to about 15 MeV, well below the threshold for
producing muons, this reaction is sensitive only to electron neutrinos.

\begin{figure}
\begin{minipage}[b]{0.5\linewidth}
\centering
\includegraphics[height=9.4cm]{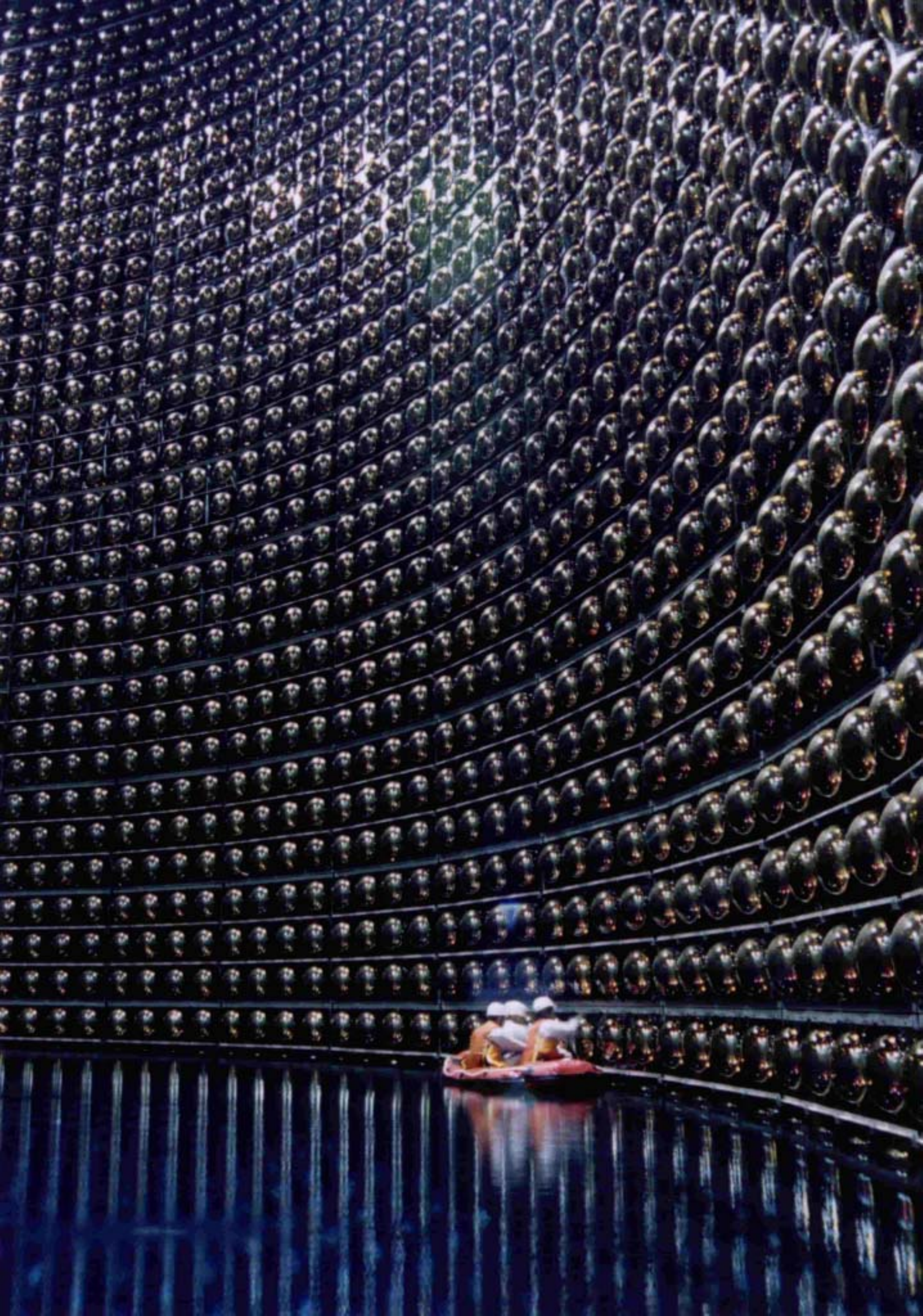}
\end{minipage}
\begin{minipage}[b]{0.5\linewidth}
\centering
\includegraphics[height=9.4cm]{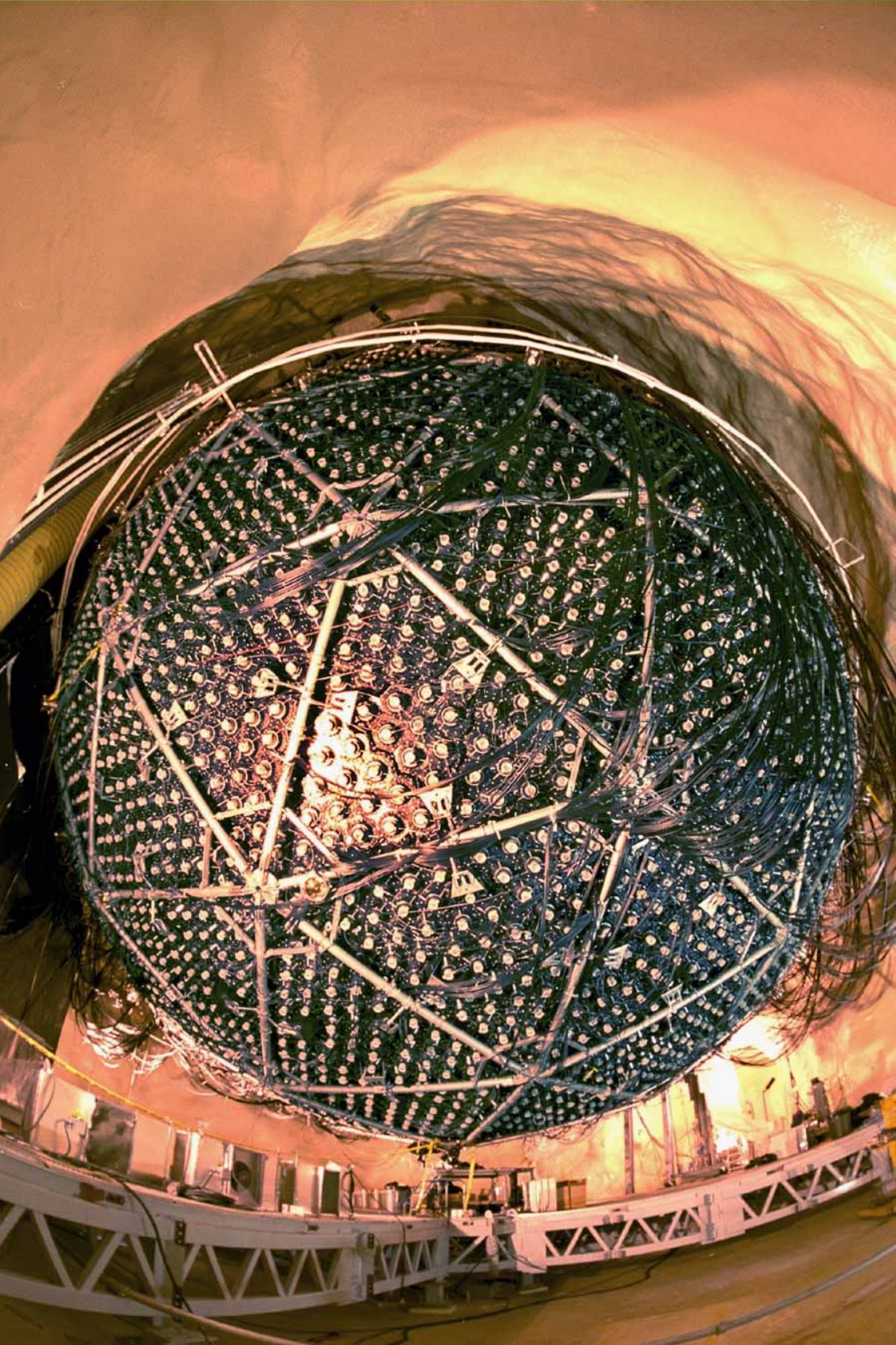}
\end{minipage}
\caption{The left panel shows the Super-Kamiokande detector during filling, with
scientists cleaning PMT surfaces as the water rises.  The right panel
is a fish-eye photo of the SNO detector and cavity, showing the PMTs
and support structure prior to cavity and detector filling.}
\label{fig:sksno}
\end{figure}

A second possible nuclear detection channel is neutral-current
scattering
\[ \nu_x + (N,Z) \rightarrow \nu_x^\prime + (N,Z)^*, \]
a process independent of the neutrino flavor.  If this scattering
leaves the nucleus in an excited state, the observable would be the
de-excitation of the nucleus, such as a decay $\gamma$ ray or the
breakup of the nucleus.  (An example will be given below, in the
discussion of SNO.)  Alternatively, neutrino elastic scattering 
(without nuclear excitation) is a coherent process at low energies,
with a cross section proportional to the square of the weak charge,
which is approximately the neutron number $N$ of the nucleus.
The signal then is the small recoil energy of the nucleus after
scattering.  

A third possibility is the scattering of neutrinos off electrons,
\[ \nu_x + e^- \rightarrow \nu_x^\prime + e^-,\]
with detection of the recoiling scattered electron.
Both electron- and heavy-flavor ($\nu_\mu$, $\nu_\tau$) solar neutrinos
can scatter off electrons, the former by charge and neutral currents,
and the latter by neutral currents only.  Consequently the cross
section for scattering heavy-flavor neutrinos is only about 0.15 that
for electron neutrinos.  This process provides a third way
of probing neutrino flavor, due to this differential
sensitivity.   An important aspect of electron-neutrino
scattering is its directionality:  for solar neutrino energies much above
the electron mass of 0.511 MeV, the electron scatters into a narrow
cone along the incident neutrino's direction.   This directionality 
provides a powerful tool for extracting solar neutrino events from 
background:  neutrino events correlate with the direction of the Sun, while
background events should be isotropic.  Thus neutrino events can be
identified as the excess seen at forward electron angles.

These various detection channels were exploited in two large-volume
water Cerenkov detectors that recorded events in real time and
provided flavor sensitivity, as well as in the liquid scintillator experiment Borexino.

Super-Kamiokande \cite{SKgen} (Fig.~\ref{fig:sksno}) is a detector consisting of 50 kilotons of ultra-pure water
within a cylindrical stainless steel tank, 39m in diameter and 42m tall.  
Two meters inside the walls a scaffold supports a dense array of 50-cm-diameter hemispherical 
photomultiplier tubes (PMTs), which face inward and view the inner 32 kilotons of
water.  Additional 20-cm tubes face outward, viewing the outer portion
of the detector that serves as a shield and as a veto.   A solar neutrino 
can interact in the inner detector, scattering off an electron.  The recoiling,
relativistic electron then produces a cone of Cherenkov radiation, a pattern
that can be reconstructed from the triggering of the phototubes that
surround the inner detector.  The detector is housed deep within Japan's
Kamioka Mine, approximately one kilometer underground.

The Super-Kamiokande detector began operations in 1996, progressing from phase I
to the current phase IV.   During Super-Kamiokande I the detector was instrumented with an 
array of 11,146 50-cm PMTs, corresponding to about 40\% coverage.    In November 2001,
during a shutdown period for repairs and upgrades,  
one of the 50-cm PMTs imploded, creating
a powerful shock wave that propagated through the tank, destroying 60\%
of the phototubes.  The detector was subsequently rebuilt  with about
half the original number of phototubes, evenly spaced over the scaffold,
so that the coverage was reduced to 20\%.  The detector operated in this
SK-II phase from late 2002 until 2005.  Following a second reconstruction
in which the phototube coverage
was restored to 40\% and other improvements made, SK-III data was obtained
from October 2006 through August 2008.   
Preliminary results for 1069 days of running in the current SK-IV phase
(in which the threshold for detecting electrons has been lowered to a (total energy) of 
4 MeV)  were 
reported in summer, 2012.   The SK-III observed event rate of scattered electrons
between 5.0 and 20.0 MeV is equivalent to an unoscillated
${}^8$B flux of (2.39 $\pm$ 0.04 (stat) $\pm$ 0.05 (sys)) $\times 10^6$/cm$^2$/s
\cite{Smy}, well below the SSM flux given in Table \ref{tab:one}.

The Sudbury Neutrino Observatory \cite{SNOgen} (SNO) (Fig.~\ref{fig:sksno}) was constructed at 
an extraordinary depth, two kilometers 
underground in the INCO Creighton nickel mine in Ontario, Canada.
The detector took data from May 1999 through November 2006, operating
in three different modes over its 7.5-year lifetime.   SNO employed a
one-kiloton target of heavy water, contained within a spherical acrylic vessel
six meters in radius.  This sphere was surrounded by an additional five meters
(seven kilotons) of very pure ordinary water, filling the rock cavity 
that housed the entire detector.  An
array of 9600 20-cm PMTs, mounted on a geodesic sphere surrounding the inner
vessel, provided 56\% coverage.  As in the case of SK-III,
SNO operated with a threshold of  5 MeV through much of its lifetime, detecting the
portion of the ${}^8$B solar neutrino spectrum from 5-15 MeV. 
(Recently a low-energy re-analysis has been completed that employed a
electron kinetic energy threshold of 3.5 MeV).

The choice of a heavy-water target allowed SNO experimentalists to 
exploit all three of the reaction channels described above, with their
varying flavor sensitivities
\begin{eqnarray}
\nu _x + e^- \rightarrow \nu_x^\prime +
e^{-}~~~&&\mathrm{ES:~elastic~scattering} \nonumber \\
\nu _e + d \rightarrow  p + p + e^{-}~~~&&\mathrm{CC:~charged~current} \nonumber \\
\nu _x + d \rightarrow  \nu_x^\prime + n +
p~~~&&\mathrm{NC:~neutral~current} \nonumber 
\end{eqnarray}
The elastic scattering (ES) reaction is the same as that employed by Super-Kamiokande, 
with its differing sensitivities to electron- and heavy-flavor neutrinos.  The charged-current (CC)
reaction on deuterium is sensitive only to electron-flavor neutrinos,
producing electrons that carry off most of 
the incident neutrino's energy (apart from the 1.44 MeV needed to
break a deuterium nucleus into p+p).   
Thus, from the energy distribution of the electrons, one can 
reconstruct the incident $\nu_e$ spectrum (and
possible distortions discussed below) more accurately than in the
case of ES.

The NC reaction, which is observed through the produced neutron,
provides no spectral information, but does measure the total solar
neutrino flux, independent of flavor. The SNO experiment has used
three techniques for measuring the neutrons. In the initial pure-D$_2$O 
phase the neutrons captured on deuterium,
producing 6.25 MeV $\gamma$s. In a second phase 2.7 tons of salt were
added to the heavy water so that Cl would be present to enhance
the capture, producing 8.6 MeV $\gamma$s. In both of these approaches
the NC and CC events can be
separated reasonably well because of the modest backward peaking
($\sim$ 1-$\cos{\theta}$/3) in the angular distribution of the
latter. This allowed the experimenters to determine the total and
electron neutrino fraction of the solar neutrino flux.  Finally, in the third
phase, direct neutron detection was provided in
pure D$_2$O by an array of ${}^3$He--filled proportional counters,
cleanly separating this signal from CC scattering.

SNO was constructed at very great depth and under clean-room 
conditions because of the need to suppress backgrounds.  In particular,
a minute amount of dust in the detector could have introduced 
environmental radioactivities that would have obscured the NC signal, a
single neutron.  The great advantage of the SNO detector was its
three distinct detection channels, sensitive to different
combinations of electron and heavy-flavor neutrinos.  Furthermore,  because the ES and
CC scattered electrons are measured in the same detector, several
important systematic effects cancel in the ratio of events.  The ES
reaction provided an important cross check on the consistency of
the results from the CC and NC channels.

\begin{figure}
\begin{center}
\includegraphics[width=12cm]{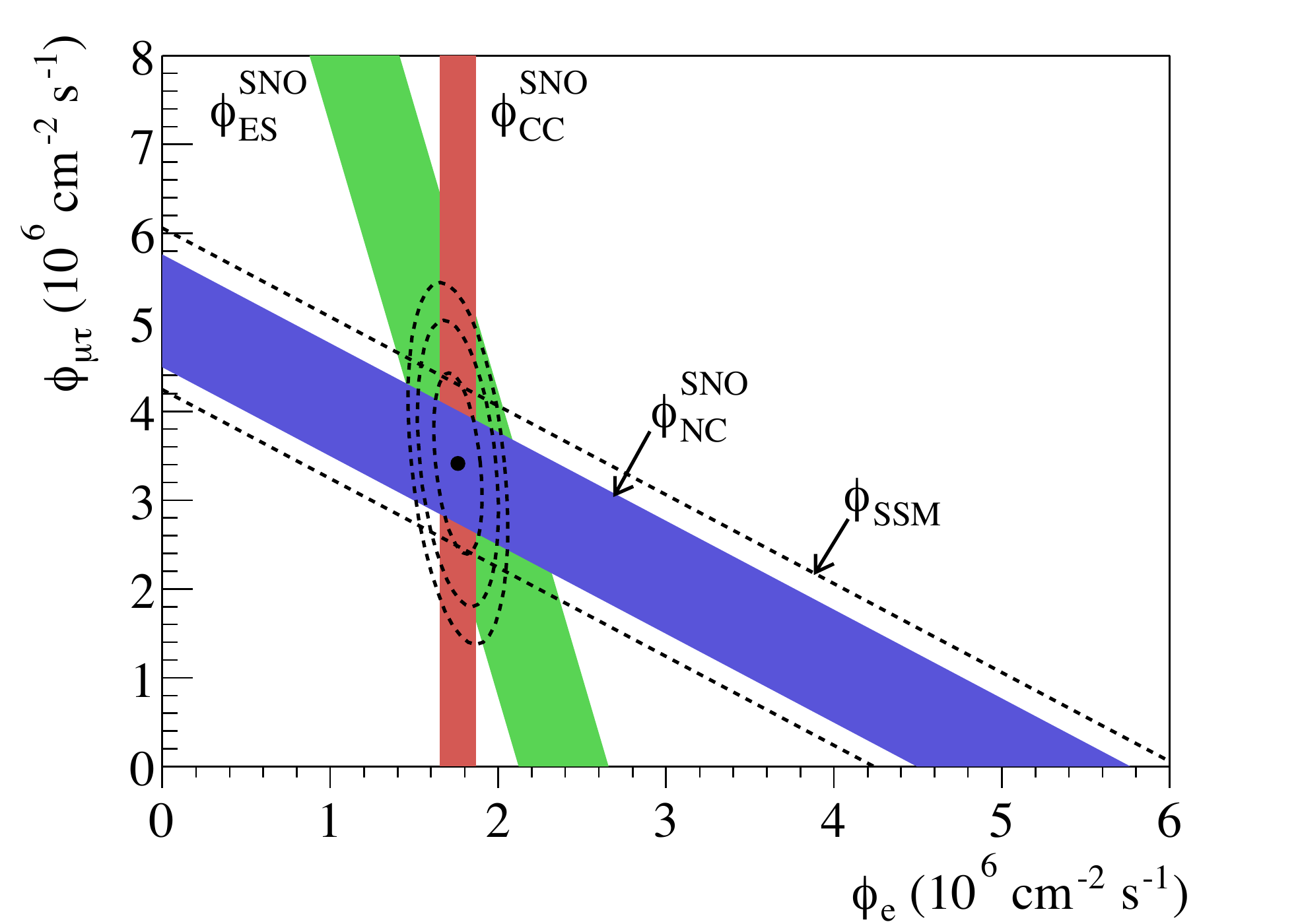}
\end{center}
\caption{ Results from the D$_2$O phase of the SNO experiment \cite{SNOgen}.  The allowed bands for
CC, NC, and ES reactions of solar neutrinos intersect to show a flux that is
one-third $\nu_e$s and two-thirds heavy-flavor
neutrinos.  There is agreement between the NC total-flux measurement and
the predictions of the SSM (band indicated by the dashed lines).  Figure courtesy of the
Sudbury Neutrino Observatory collaboration.}
\label{fig:SNOresults}
\end{figure}

The results from the three phases of SNO are in generally good agreement,
separately and in combination establishing a total flux of active neutrinos from
${}^8$B decay of $\phi_\mathrm{NC}(\nu_\mathrm{active}) = (5.25 
\pm 0.16 (\mathrm{stat})^{+0.11}_{-0.13} (\mathrm{syst})) \times 10^6$/cm$^2$/s, in good agreement
with SSM predictions.  SNO also established $\phi_\mathrm{CC}(\nu_e) \sim 0.34
\phi_\mathrm{NC}(\nu_\mathrm{active})$.  Thus, as
Fig.~\ref{fig:SNOresults} illustrates,
about two-thirds of the electron
neutrinos produced in the Sun arrive on Earth as heavy-flavor
(muon or tauon) neutrinos.   The Davis detector and the
CC channel in SNO are blind to these heavy flavors, seeing only
the portion with electron flavor.  Thus the solar neutrino problem
was not a matter of missing neutrinos, but rather one of neutrinos
in hiding.  The implications of this discovery -- that neutrinos are
massive and violate flavor -- are profound, indicating that our
Standard Model of particle physics is incomplete.

Other potential signals of neutrino oscillations in matter, such as an energy-dependent
distortion in the $\nu_e$ survival probability or day-night differences due to neutrino
passage through the earth, were not seen in SNO, nor have they emerged from
SK analyses to date at a convincing level of confidence.

\begin{figure}
\begin{center}
\includegraphics[width=12cm]{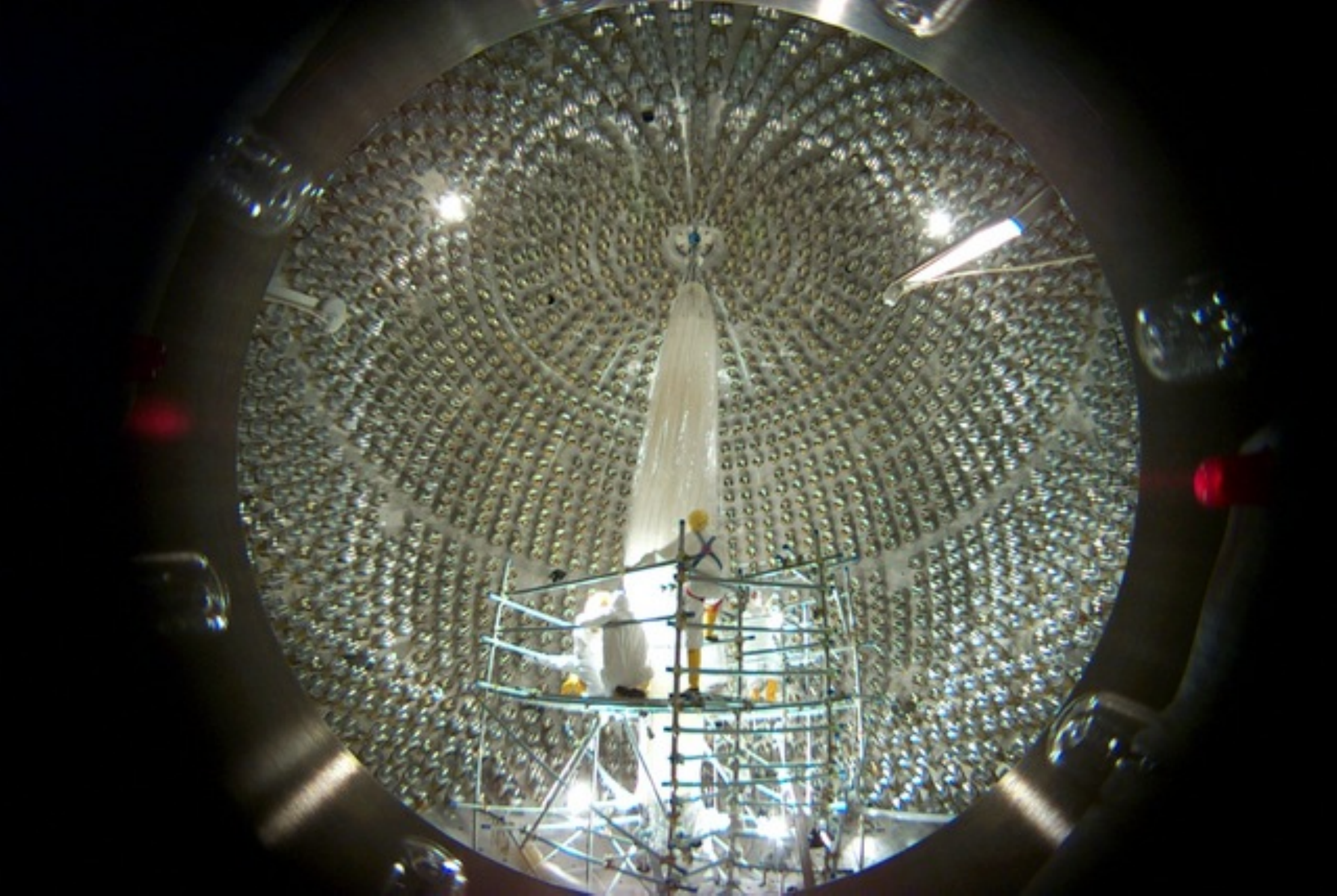}
\end{center}
\caption{ The Borexino inner vessel during installation.}
\label{fig:Borexino}
\end{figure}

The Borexino experiment \cite{Borgen}  (Fig.~\ref{fig:Borexino}), located in Italy's Gran Sasso Laboratory, is the
first to measure low-energy ($\lsim$ 1 MeV) solar neutrinos in real time.   The
detector is housed within a 16.9m domed tank containing an outer layer of ultrapure water that provides
shielding against external neutrons and gamma rays.  At the inner edge of the water a stainless steel 
sphere serves as a support structure for an array of photomultiplier tubes that view both the inner
detector and the outer water shield, so that the Cherenkov light emitted by muons passing through
the water can be used to veto those events.  Within the steel sphere there are two regions, separated
by thin nylon vessels, containing high-purity buffer liquid, within which is sequestered a central volume
of 278 tons of organic scintillator.  The fiducial volume consists of $\sim$ 100 tons of the
liquid scintillator at the very center of the detector.  Scintillation light produced by recoil
electrons after ES events is the solar neutrino signal.  The 862 keV $^7$Be 
neutrinos produce a recoil electron spectrum with a distinctive cut-off edge 
at 665 keV.

Results reported by the  Borexino Collaboration from 2008 through 2012 \cite{Borgen}
constrain three low-energy
solar neutrino branches.  The Collaboration
\begin{enumerate}
\item found a $^7$Be solar rate equivalent to an unoscillated flux of
$(3.10 \pm 0.15) \times 10^9$/cm$^2$s, or about 62\% of the GS98-SFII SSM central value;
\item made the first direct, exclusive determination of the pep flux,
$(1.6 \pm 0.3) \times 10^8$/cm$^2$s (95\% c.l.); and
\item established a limit on the CNO neutrino flux, $\phi_\mathrm{CNO} < 7.7 \times 10^8$/cm$^2$s
at 95\% c.l.
\end{enumerate}

\subsection{Neutrino Mass and Oscillations}
The phenomenon by which a massive neutrino of one flavor
changes into one of a second flavor is called neutrino oscillations.  
Neutrino oscillations have been shown to be responsible not
only for the missing solar neutrinos in Davis's experiment, but also
for the missing atmospheric neutrinos that will be discussed
in the next chapter.  Neutrino
oscillations can be altered by the presence of matter or the
presence of other neutrinos.  For this reason, astrophysical ``laboratories" for
studying neutrinos -- the Sun, supernovae, the early universe -- are
of great interest because of the unique conditions they
provide, including very
long ``baselines" over which neutrino propagate and enormous matter
densities and neutrino fluences.

Neutrino oscillations originate from two distinct sets of labels
carried by neutrinos.  One is flavor, a property of the 
weak interaction: an electron neutrino is defined as the neutrino
accompanying a positron in $\beta$ decay.  The other possible label
is mass.  If a neutrino
has a mass $m$, it propagates through free 
space with an energy and three momentum related by
$\omega = \sqrt{\vec{k}^2 + m^2}$.    Thus neutrino states
can be labeled according to flavor, and also labeled according
to their masses.

However nothing requires the neutrinos of definite flavor to
be coincident with the neutrinos of definite mass.  (In fact, in the
analogous case of the quarks, it has long been known that the
flavor (or weak interaction) eigenstates are not identical to the
mass eigenstates: That is, the up
quark decays not only to the down quark, but also occasionally
to the strange quark.)  Neutrino oscillations occur when
the mass eigenstates $|\nu_1 \rangle$ and $|\nu_2 \rangle$ (with masses
$m_1$ and $m_2$) are related to the weak interaction eigenstates by
\begin{eqnarray}
|\nu_e\rangle  &=& \cos \theta_v |\nu_1\rangle  + \sin \theta_v|\nu_2 \rangle  \nonumber \\
|\nu_\mu\rangle &=& - \sin \theta_v |\nu_1 \rangle + \cos \theta_v |\nu_2 
\rangle \nonumber
\end{eqnarray}
where $\theta_v$, the (vacuum) mixing angle, is nonzero.  (Here, for
simplicity, we consider just two neutrinos -- the generalization to three
flavors will be described later.)
  
\begin{figure}
\begin{center}
\includegraphics[width=12cm]{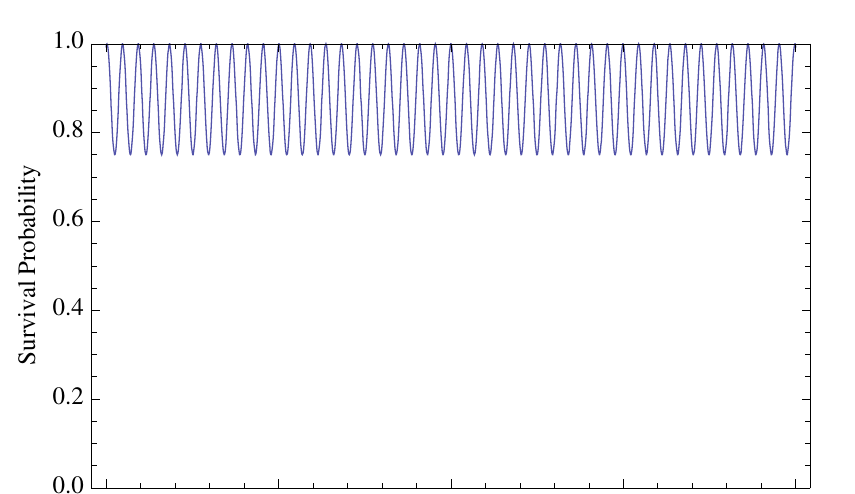}
\includegraphics[width=12cm]{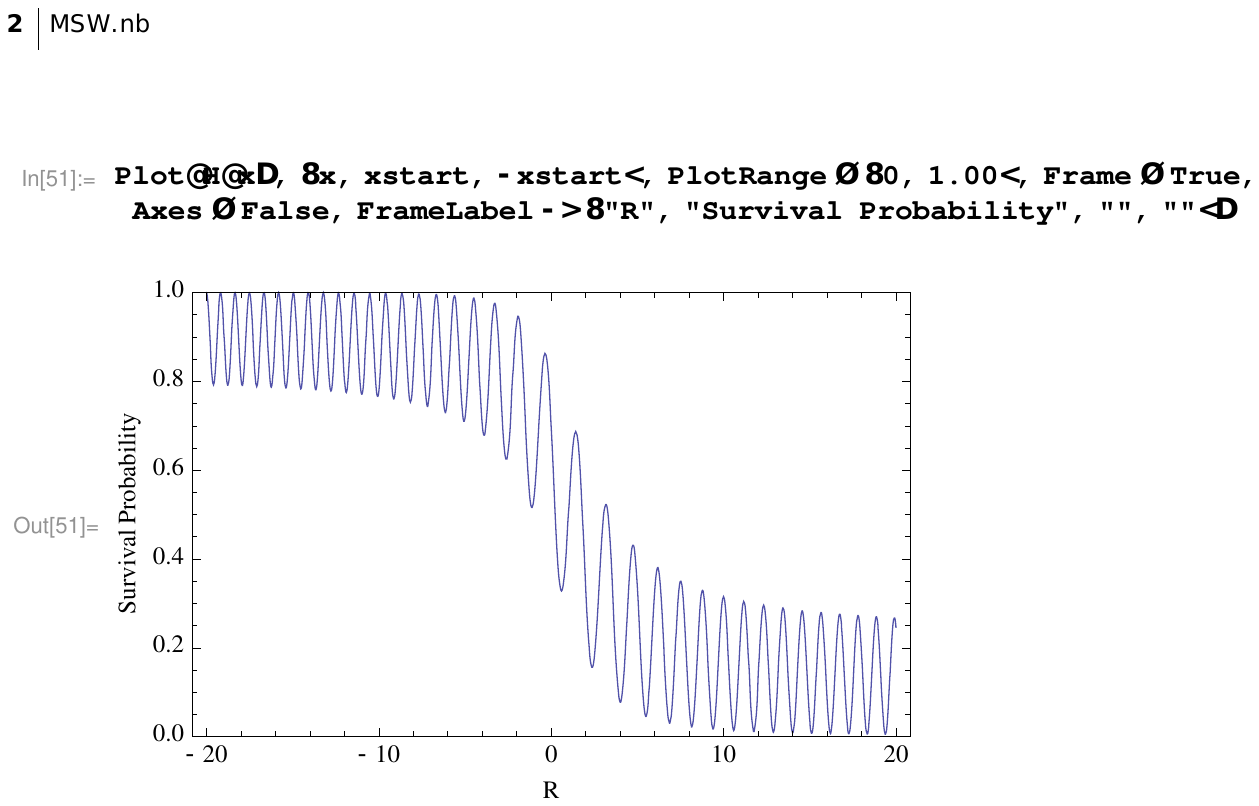}
\end{center}
\caption{A simple example illustrating the MSW mechanism.  The top frame
shows vacuum oscillations for a $\nu_e$ created at $R=-20$ and propagating
to the right, for $\theta_v$=15$^\circ$.  The average $\nu_e$ survival
probability is large, 87.5\%.  (Here
the distance R is given in units related to the oscillation length,
$4 E \cos{2 \theta}/(\delta m^2 \sin^2{2 \theta}$).)  In the bottom frame an electron
density $\rho(R)$ has been added proportional to $1- (2/\pi) \tan^{-1}{a R}$, with $a$ chosen to
guarantee adiabaticity, and normalized so that  1) $\rho(r) \rightarrow 0$ as $R \rightarrow \infty$;
2) the matter effects cancel the vacuum mass difference for $R \sim 0$ (the MSW crossing point);
and 3) the matter effects reverse the sign of $\delta m^2$ as $R \rightarrow -\infty$.  Thus these are the
MSW conditions described in the text.  A $\nu_e$ created 
at high density ($R =-20$), where it approximately coincides with the local heavy-mass
eigenstate, adiabatically propagates to low-density ($R=+20$), where it approximately
coincides with the $\nu_\mu$.   Thus the $\nu_e$
survival probability at $R=20$ is much reduced, compared to the vacuum case.  Note
that the local oscillation length is maximal near the crossing point.}
\label{fig:osc}
\end{figure}

In this case a state produced as a $|\nu_e\rangle$
or a $|\nu_\mu\rangle$ at some time $t$ --- for example, a neutrino
produced by $\beta$ decay in the Sun's core --- does not remain a pure flavor eigenstate
as it propagates away from the source.  The different
mass eigenstates comprising the neutrino will accumulate different
phases as the neutrino propagate downstream, a phenomenon known as 
vacuum oscillations (vacuum because the experiment is done in free
space).  While at time $t$=0 the neutrino is a flavor eigenstate
\[ |\nu(t=0)\rangle  = |\nu_e \rangle = \cos \theta_v |\nu_1\rangle  + \sin \theta_v|\nu_2 \rangle ,\]
the accumulate phases depend on the mass
\[e^{i(\vec{k} \cdot \vec{x} - \omega t)} =
e^{i [ \vec{k} \cdot \vec{x} - \sqrt{m_i^2 + k^2}~t ]} . \]
If the neutrino mass is small compared to the neutrino 
momentum/energy, one finds
\[ |\nu(t) \rangle = e^{i(\vec{k} \cdot \vec{x} - kt
-(m_1^2+m_2^2)t/4k)} \left( \cos \theta_v |\nu_1 \rangle e^{i \delta m^2 t/4k}
+ \sin \theta_v |\nu_2 \rangle e^{-i \delta m^2 t/4k} \right) \] 
There is a common average phase (which has no physical
consequence) as well as a beat phase that depends on
\[ \delta m^2 = m_2^2 - m_1^2.  \]
From this one can find the probability that 
the neutrino state remains a $|\nu_e\rangle$ at time t
\[ P_{\nu_e} (t) = | \langle \nu_e | \nu(t) \rangle |^2  =
1 - \sin^2 2 \theta_v \sin^2 \left({\delta m^2c^4 x\over 4 \hbar c E} \right)  \]
The probability oscillates from 1 to  $1-\sin^2 2 \theta_v$ and back to 1
over an oscillation length scale
\[ L_\mathrm{o} = {4 \pi \hbar c E \over \delta m^2 c^4} ,\]
as depicted in the upper panel of Fig.~\ref{fig:osc}.
In the case of solar neutrinos, if $L_\mathrm{o}$ were comparable to or shorter than one astronomical unit, a 
reduction in the solar $\nu_e$ flux would be expected in terrestrial detectors.
  
The suggestion that neutrinos could oscillate was first made by
Pontecorvo in 1958, who pointed out the analogy with $K_0 \leftrightarrow \bar 
K_0$     
oscillations.  If the Earth-Sun separation is much larger than $L_\mathrm{o}$, one expects 
an average flux reduction due to oscillations of
\[ 1 - {1 \over 2} \sin^2 2 \theta_v . \]
For a 1 MeV neutrino, this requires $\delta m^2c^4 \gg 10^{-12}$ eV$^2$.  
But such a reduction $-$ particularly given the initial theory prejudice that neutrino mixing
angles might be small $-$ did not seem sufficient
to account for the factor-of-three discrepancy that emerged from Davis's early
measurements.

The view of neutrino oscillations changed 
when Mikheyev and Smirnov \cite{MS} showed in 1985 that neutrino
oscillations occurring in matter -- rather than in vacuum -- could
produce greatly enhanced oscillation probabilities.   This
enhancement comes about because neutrinos propagating through matter 
acquire an additional mass due to their interactions with the
matter.  In particular, because the Sun contains many electrons, 
the electron neutrino becomes heavier in proportion to the
local density of electrons.   An enhanced probability for oscillations can
result when an electron neutrino passes from a high-density region
(such as the solar core) to a low-density one (such as the surface
of the Earth).  This matter enhancement is
called the MSW mechanism after
Mikheyev, Smirnov, and Wolfenstein \cite{Wolf} (who first described the 
phenomenon of neutrino effective masses).

To explain this enhancement, consider the case where the vacuum mixing
angle $\theta_v$ is small  and $m_2 > m_1$.
Then in vacuum $|\nu_e\rangle  \sim  |\nu_1\rangle \equiv |\nu_L(\rho=0) \rangle$
where $\rho=0$ is the electron density in vacuum, that is,
the $\nu_e$ and the light vacuum eigenstate $|\nu_L(\rho=0) \rangle$ are almost identical.
(Correspondingly, the heavy eigenstate $|\nu_2 \rangle \equiv |\nu_H(\rho=0) \rangle
 \sim |\nu_\mu \rangle$ in vacuum.)
Now what happens in matter?  As matter makes the $\nu_e$
heavier in proportion to the electron density, if that density
is sufficiently high, clearly the electron neutrino
must become the (local) heavy mass eigenstate.  That is, 
$|\nu_e\rangle  \sim |\nu_H(\rho \rightarrow \infty) \rangle$ (and
consequently $|\nu_\mu\rangle  \sim  |\nu_L(\rho \rightarrow \infty) \rangle$).
That is, we conclude that there must be a local mixing angle $\theta(\rho)$
that rotates from $\theta_v \sim 0$
in vacuum to $\theta(\rho) \sim \pi/2$ as $\rho \rightarrow \infty$. 

MSW enhancement occurs when the density changes between
neutrino production and detection.    In particular, electron neutrinos
produced in the high-density solar core are created as heavy mass
eigenstates.  If these neutrinos now propagate to the solar surface
adiabatically  -- this means that changes in the solar density scale height
$\rho^{-1}~ d \rho/dx$ are 
small over an oscillation length, at all points along
the neutrino trajectory -- they will remain on the heavy-mass trajectory,
and thus exit the Sun as $|\nu_H(\rho=0)\rangle =
|\nu_2 \rangle \sim |\nu_\mu \rangle$. That is, there
will be an almost complete conversion of the $\nu_e$s produced
in the solar core to $\nu_\mu$s.  The MSW mechanism is an example
of an avoided level crossing, a familiar phenomenon in quantum
mechanics.

A schematic comparison of vacuum (upper panel) and matter-enhanced (lower panel) oscillations
is shown in Fig.~\ref{fig:osc}.  The matter transition between electron
and muon flavors is centered around a density where the vacuum mass 
difference is just compensated by the matter contributions.

The discussion above was presented for two neutrino flavors, and thus a single
vacuum mixing angle $\theta_v$ and mass difference $\delta m^2$.  But three neutrino
flavors appear in the Standard Model of particle physics.  In this case the
relationship between flavor \{$\nu_e,\nu_\mu,\nu_\tau$ \} and mass \{ $\nu_1,\nu_2,\nu_3$ \} eigenstates
is described by the  PMNS matrix \cite{MNS,Pontecorvo67}
\begin{equation}
\left( \begin{array}{c} | \nu_e \rangle \\ | \nu_\mu \rangle \\ | \nu_\tau \rangle \end{array} \right) =
\left( \begin{array}{ccc} c_{12} c_{13} & s_{12} c_{13} & s_{13} e^{-i \delta} \\
-s_{12}c_{23}-c_{12}s_{23} s_{13} e^{i \delta} & c_{12}c_{23}-s_{12}s_{23}s_{13} e^{i \delta} & s_{23} c_{13} \\
s_{12}s_{23}-c_{12}c_{23}s_{13} e^{i \delta} & -c_{12}s_{23}-s_{12}c_{23}s_{13}e^{i \delta} & c_{23} c_{13} \end{array}
\right) \left( \begin{array}{c} e^{i \alpha_1 /2} | \nu_1\rangle \\ e^{i \alpha_2 /2} | \nu_2 \rangle \\ | \nu_3 \rangle \end{array} \right)
\end{equation}
where $c_{ij} \equiv \cos{\theta_{ij}}$ and $s_{ij} \equiv \sin{\theta_{ij}}$.  This matrix depends on
three mixing angles $\theta_{12}$, $\theta_{13}$, and $\theta_{23}$, of which the first and last are the
dominant angles for solar and atmospheric oscillations, respectively; a Dirac phase $\delta$ that
can induce CP-violating differences in the oscillation probabilities for
conjugate channels such as $\nu_\mu \rightarrow \nu_e$ versus
$\bar{\nu}_\mu \rightarrow \bar{\nu}_e$; and two Majorana phases $\alpha_1$ and $\alpha_2$
that will affect the interference among mass eigenstates in the effective neutrino mass probed
in the lepton-number-violating process of neutrinoless double $\beta$ decay.  There are
also two independent mass$^2$ differences, $\delta m_{21}^2 \equiv m_2^2-m_1^2$ and
$\delta m_{32}^2 \equiv m_3^2-m_2^2$.

In this framework, the dominant oscillation affecting solar neutrinos is that described
by $\delta m_{21}^2$ and $\theta_{12}$.
The results from SNO, Super-Kamiokande, Borexino, and earlier solar
neutrino experiments, and from the
reactor experiment KamLAND \cite{KamLAND}, have determined these parameters
quite precisely, as discussed below.   Unlike the MSW
example given above, the dominant mixing is characterized by a  large mixing angle,
$\theta_{12} \sim 34^\circ$.  Thus the vacuum oscillation probability is significant.
The mass$^2$ difference,
$\delta m_{21}^2 \sim 7.6 \times 10^{-5}$ eV$^2$, leads to important
matter effects in the higher energy portion of the solar neutrino
spectrum, thus influencing the rates found in the SNO, Super-Kamiokande,
and chlorine experiments.  These effects produce a characteristic,
energy-dependent distortion of the solar $\nu_e$ spectrum.

\subsection{Solar Neutrinos: Oscillation Parameters and Outlook}
Neutrino oscillations proved to be responsible for both the solar
neutrino problem and the atmospheric neutrino problem discussed
in the next chapter.
This phenomenon requires both flavor mixing and neutrino mass,
phenomena that can be accommodated in various extensions of the
Standard Model of particle physics.  The great interest in neutrino astrophysics
stems in part from the expectation that newly discovered neutrino properties
may help us formulate the correct extension.  Indeed, the tiny neutrino
mass differences deduced from the solar and atmospheric problems are
compatible with theories where the neutrino mass is inversely 
proportional to a scale for new physics of about 10$^{15}$ GeV.  This is
close to the Grand Unified scale where supersymmetric extensions
of the Standard Model predict that the strengths of the fundamental forces
unify.

The precision with which fundamental parameters of the Standard Model of
particle physics were determined by solar neutrino experiments, supplemented
by KamLAND's measurements to better constrain $\delta m_{21}^2$, is really
quite remarkable.  The global analysis performed by the SNO group of this
set of experiments yielded
\[ \sin^2{\theta_{12}} = 0.308 \pm 0.014~~~~~~~\delta m_{21}^2 =(7.41^{+0.21}_{-0.19}) \times 10^{-5}
\mathrm{eV^2}~~~~~~~\sin^2{\theta_{13}}=0.025^{+0.018}_{-0.015} \]
where $\theta_{13}$ is the mixing angle for the subdominant 1-3 oscillation.
The first two results are in excellent agreement, in value and uncertainty, with the
corresponding values from the Bari \cite{Bari} and Valencia \cite{Valencia}  global analyses that include all
accelerator and reactor data.  The main impact of terrestrial neutrino experiments on solar neutrino analyses has
come from the Daya Bay, RENO, and Double Chooz reactor experiments, which have
significantly tightened the constraints on $\theta_{13}$,
\[ \sin^2{\theta_{13}} = \left\{ \begin{array}{lr} 0.0243^{+0.0027}_{-0.0026} & ~~~\mathrm{Bari} \\
0.0248^{+0.0031}_{-0.0029} & ~~~\mathrm{Valencia}  \end{array} \right.  \]
where the error bars include the uncertainty in the mass hierarchy (that is,
the sign of $\delta m_{32}^2$).  Yet even in this case
the solar neutrino analysis gave the correct central value for $\theta_{13}$.

The effects of matter on neutrino oscillations have also been convincingly demonstrated from
the comparison of the Borexino result for ${}^7$Be neutrinos and those of SNO and Super-Kamiokande
for ${}^8$B neutrinos.  The ${}^7$Be neutrinos lie in the vacuum oscillation region where the survival
probability is larger; the SNO and Super-Kamiokande detectors are sensitive to more energetic
${}^8$B neutrinos,
where matter effects enhance the oscillation into heavy-flavor states.

There are also interesting developments involving the SSM, as we have noted 
in our discussion of two competing SSMs, GS98-SFII and AGSS09-SFII.
One of the important validations of the SSM has come through helioseismology,
the measurement of solar surface fluctuations, as deduced from the Doppler
shifts of spectral lines.  The observed patterns can be inverted to determine
properties of solar interior, including the sound speed
as a function of the solar radius.  For about a decade the agreement between SSM
predictions and observation had been excellent.  

But recently improved three-dimensional models of the photosphere, when applied
to the analysis of absorption lines,
lead to a $\sim$ 30\% downward revision in convective-zone metal abundances 
(and thus Z$_\mathrm{ini}$).   If the new abundances are employed in the SSM, the resulting
changes in the opacity alter
both neutrino flux predictions and the sound speed profile. 
This solar abundance problem could indicate that the SSM assumption of a
homogeneous zero-age Sun -- which is not obviously correct due to the efficiency
with which planet formation swept metals from the proto-solar disk --  may have to be
re-examined.  One of the goals of the next solar neutrino experiment -- SNO+, a larger,
deeper version of Borexino under construction in the cavity that formerly held SNO --
is to use CN neutrinos to measure the core abundance of carbon and nitrogen
directly, thereby allowing a direct comparison between core and surface
metal abundances \cite{HRS}.

\section{Atmospheric Neutrinos}
\noindent
The atmospheric neutrino problem developed very much in parallel with
the solar neutrino problem and also involved missing neutrinos.   The first
definitive claim that neutrinos are massive came the atmospheric neutrino group
associated with Super-Kamiokande, in 1998.   The oscillations seen in atmospheric
neutrinos differ from those seen in solar neutrinos, resulting from the coupling
of a different pair of neutrinos.

\subsection{The Neutrino Source}When primary cosmic-ray protons and nuclei hit the upper atmosphere, the ensuing
nuclear reactions with atmospheric oxygen and nitrogen nuclei produce secondaries
such as pions, kaons, and muons.    Atmospheric neutrinos arise from the decay of 
these secondaries.   For energies less
than $\sim$ 1 GeV, the secondaries decay prior to reaching the Earth's surface
\begin{eqnarray}
\pi^{\pm} (K^{\pm}) &\rightarrow &\mu^{\pm} + \nu_{\mu}
(\overline{\nu}_{\mu}), \nonumber\\ \mu^{\pm} &\rightarrow & e^{\pm} +
\nu_e (\overline{\nu}_e) +  \overline{\nu}_{\mu} (\nu_{\mu}).
\end{eqnarray}
Consequently one expects the ratio
\begin{equation}
r = (\nu_e + \overline{\nu}_e) / (\nu_{\mu} + \overline{\nu}_{\mu})
\end{equation}
to be approximately 0.5 in this energy range. Detailed Monte Carlo
calculations, including the effects of muon
polarization, give $  r \sim 0.45$.  This ratio should be rather insensitive to
theoretical uncertainties.   It does not depend on absolute fluxes, and as
a ratio of related processes, one expects many sources of systematic error 
to cancel.  Indeed, while various groups have estimated  this ratio, sometimes
starting with neutrino fluxes which vary in magnitude by up to
25\%, agreement in the ratio has been found at the level of a few percent. 
This agreement persists at higher energies, where $r$ decreases because
higher energy muons survive passage through the atmosphere, due to the
effects of time dilation.

Atmospheric neutrinos are a very attractive astrophysical source for 
experimenters.  Apart from relatively minor geomagnetic effects, atmospheric 
neutrino production is uniform over the Earth.  Thus an experimenter, operating
an underground detector at some location, can make use of a set of nearly equivalent
neutrino sources at distances ranging from 10s of kilometers (directly overhead) to 13,000
kilometers (directly below, produced on the opposite side of the Earth).
Effects such as neutrino oscillations, which depend on the distance from the source 
to the target, might show up as a characteristic dependence of neutrino flux on
the zenith angle, provided the relevant
oscillation length is comparable to or less than the Earth's diameter.
(Note that the solar neutrino $\delta m_{21}^2$, for atmospheric
neutrinos of energy $\sim$ 1 GeV, would not satisfy this condition, as the
oscillation length is several times the Earth's diameter.)

\subsection{Atmospheric Neutrinos and Proton Decay Detectors}
The atmospheric neutrino anomaly grew out of efforts to build large underground detectors for
proton decay, one of the phenomena expected in the Grand Unified Theories that
were formulated in the late 1970s and early 1980s.  As atmospheric neutrinos and proton
decay would deposit very similar energies in such detectors, studies of atmospheric
neutrinos were a natural second use of such detectors.   Significant indications
of an anomaly came from the IMB \cite{IMB} and Kamiokande \cite{Kamioka} proton decay detectors.  IMB first noticed a possible deficit of neutrino-induced muon events in 1986, while
Kamiokande established a deficit in excess of 4$\sigma$ by 1988.   By 1998
this anomaly was also apparent in data from the Soudan detector and from Super-Kamiokande.

The quantity determined in such experiments is a ratio 
(observed to predicted) of ratios
\[ R = {(\nu_{\mu} / \nu_e)_{\rm data} \over (\nu_{\mu} / \nu_e)_{\rm
Monte  Carlo} } \]
where the numerator is determined experimentally, and the denominator
calculated.   Agreement between data and theory thus requires $R \sim1$.
Early experimenters faced a difficulty in evaluating this ratio
due to limited statistics:  the counting rates were too low to allow a detailed 
analysis based on the zenith angle, that is, based on the neutrino path length.
This changed with the construction of Super-Kamiokande, which provided a fiducial
volume of about 20 ktons.  An early analysis from Super-Kamiokande found
\[  R=0.61 \pm 0.03 ({\rm stat}) \pm 0.05 ({\rm syst}) \]
for sub-GeV events which were fully contained in the detector and
\[ R=0.66 \pm 0.05 ({\rm stat}) \pm 0.08 ({\rm syst}) \]
for fully- and partially-contained multi-GeV events.   In addition, the 
collaboration presented an analysis in 1998, based on 33 kton-years of data,
showing a zenith angle dependence inconsistent with theoretical calculations
of the atmospheric flux, in the absence of oscillations \cite{SKatmos}.  This indicated
a distance dependence in the muon deficit, a signature of oscillations.  
Furthermore the parameters of the oscillation, particularly  $5 \times 10^{-4}$ eV$^2 <
|\delta m_{32}^2| < 6 \times 10^{-3}$ eV$^2$, differed from those that would
later be determined from solar neutrinos.  The collaboration concluded
that the data were consistent with the two-flavor oscillation $\nu_\mu \rightarrow
\nu_\tau$.  This was the first definitive claim for massive neutrinos.

\begin{figure}
\begin{center}
\includegraphics[width=12cm]{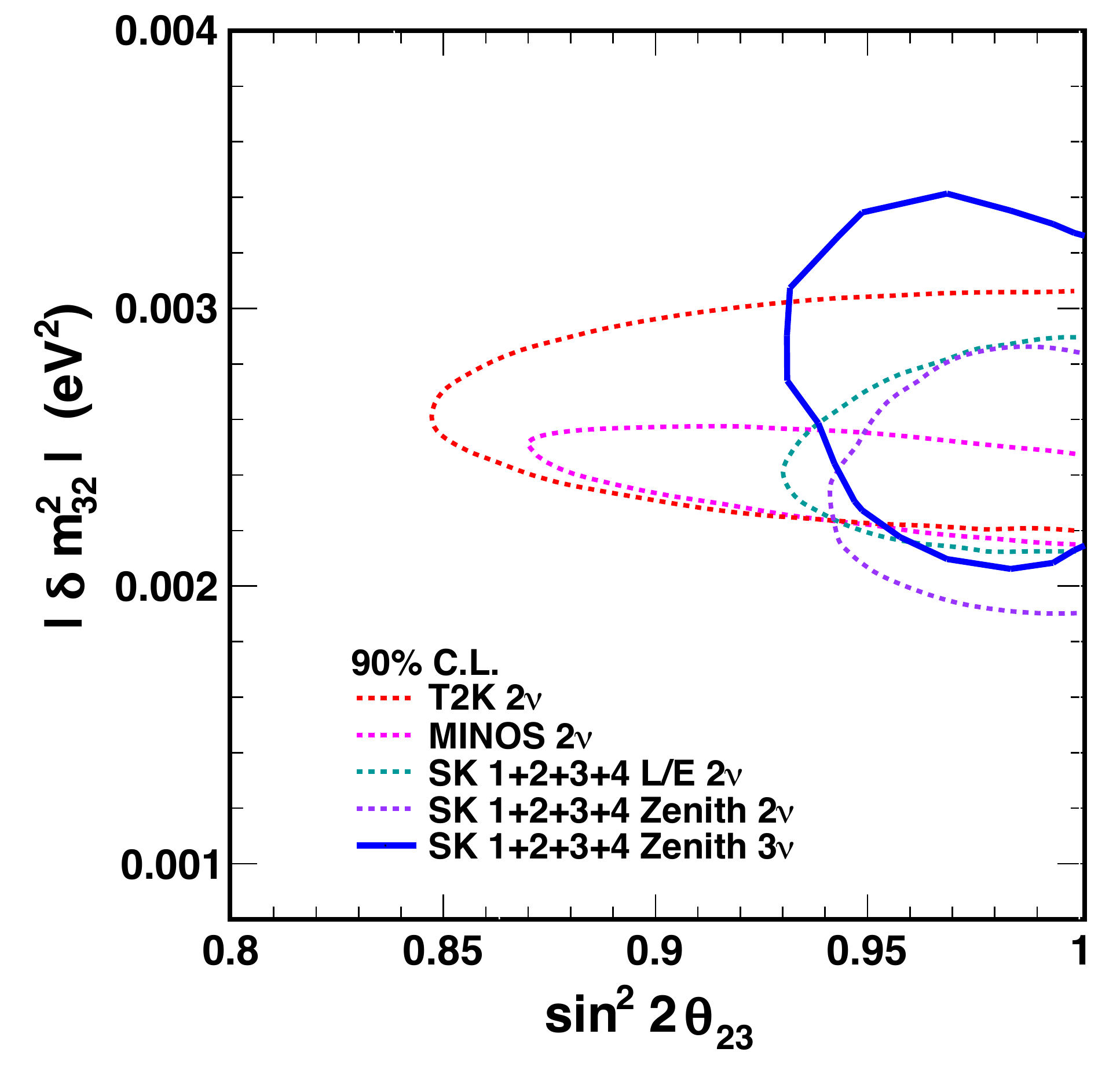}
\end{center}
\caption{ The Super-Kamiokande I-IV analysis of atmospheric neutrino oscillations, including
comparisons to the long-baseline accelerator results of MINOS and T2K.  
From \cite{Itow}, with permission of the Super-Kamiokande Collaboration.}
\label{fig:atmospheric}
\end{figure}

SK-I collected approximately 15,000 atmospheric neutrino events
in nearly five years of running.  
The collaboration's zenith-angle analysis of the data found
evidence of a first oscillation minimum at $L/E \sim 500$ km/GeV, so that
$L_\mathrm{o} \sim$ 1000 km for a 1 GeV muon neutrino.   Current results are based
on 3903 days of data from SKI-IV, including 1097 days from the current phase IV.
The entire data set has been reanalyzed with a common set of improved tools,
in a three-flavor analysis in which $\sin^2{\theta_{13}}$ was fixed to 0.025, based
on the recent reactor neutrino results from the Daya Bay, RENO, and Double Chooz experiments.
The analysis yielded \cite{Itow}
\[  \mathrm{Normal~hierarchy:}~~~~~   \begin{array}{lr} ~~~\delta m_{32}^2 = & (2.66^{+0.15}_{-0.40}) \times 10^{-3}~\mathrm{eV}^2~~~~~~(1\sigma) \\
\sin^2{\theta_{23}} = & 0.425^{+0.194}_{-0.034}  ~~~~~~(90\% \mathrm{c.l.}) \end{array} \]
\[\mathrm{Inverted~hierarchy:} ~~  \begin{array}{lr} ~~~\delta m_{32}^2 = & -(2.66^{+0.17}_{-0.23}) \times 10^{-3}~~\mathrm{eV}^2~~~~~~(1\sigma) \\
\sin^2{\theta_{23}} = & 0.575^{+0.055}_{-0.182} ~~~~~~(90\% \mathrm{c.l.}) \end{array} \]
The results continue to allow maximal mixing, $\sin^2{\theta_{23}} \sim 0.5$.  The significant difference
between the normal and inverted hierarchy best values for $\sin^2{\theta_{23}}$ reflects the
flatness of the $\chi^2$ fit around $\sin^2{\theta_{23}} \sim 0.5$, with slight local minima
appearing above and below this value.
The atmospheric value for $\delta m_{32}^2$ is consistent with the somewhat more
precise values from the MINOS (and T2K) long-baseline experiments.  The MINOS beam-neutrino analysis,
which includes atmospheric neutrino data obtained with that detector, determined
$|\delta m_{32}^2| = 2.39^{+0.09}_{-0.10} \times 10^{-3}$ eV \cite{MINOS}, a number consistent
with that from the SK atmospheric analysis.

\subsection{Outlook}
\begin{figure}
\begin{center}
\includegraphics[width=12cm]{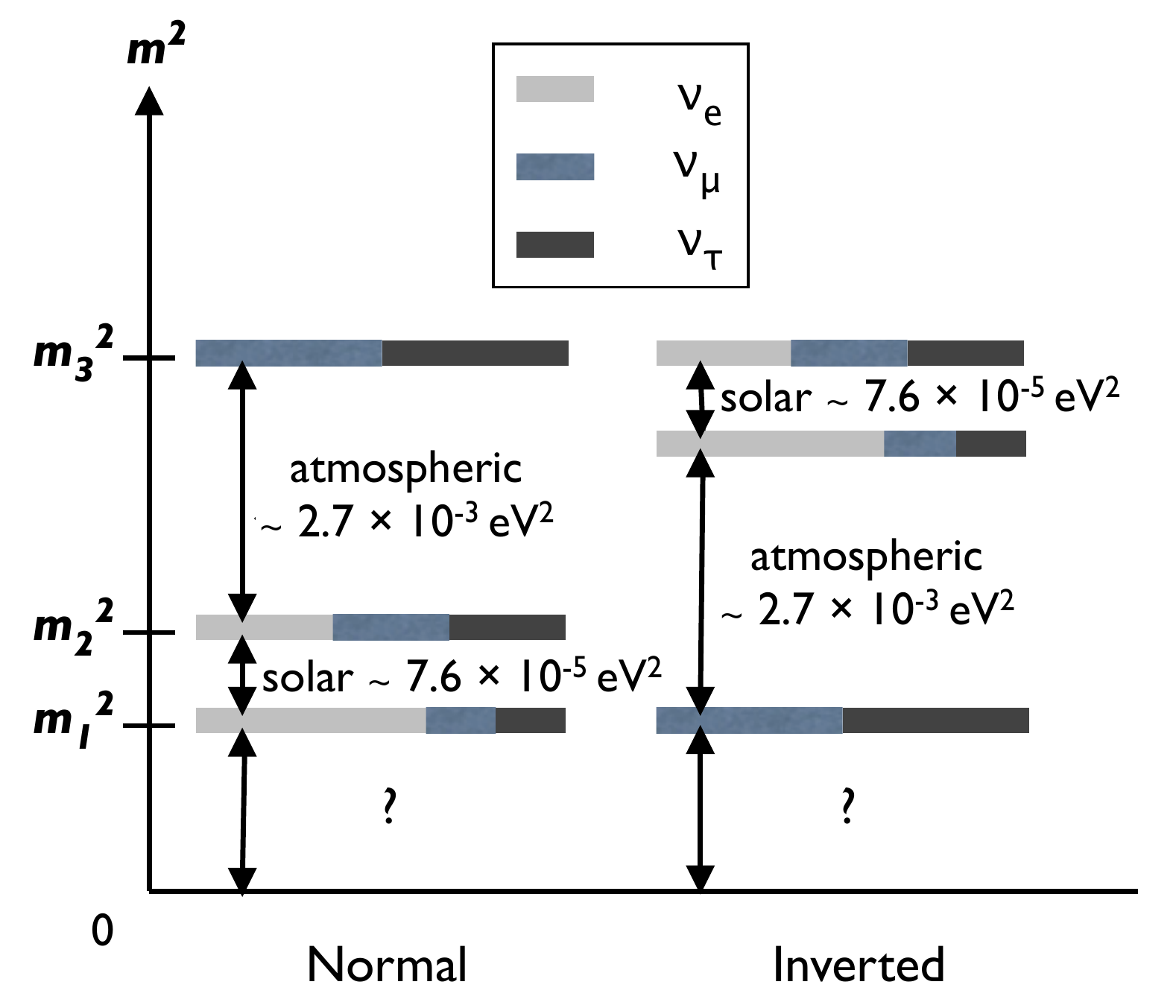}
\end{center}
\caption{ Illustration of the two level schemes that are possible for the light
neutrinos, given that no matter effects have yet been seen in atmospheric neutrinos.}
\label{fig:hierarchy}
\end{figure}

While a great deal of new physics has been learned from experiments on atmospheric
and solar neutrinos, several important questions remain \cite{APS}:
\begin{itemize}
\item Oscillation experiments are sensitive to differences in the squared masses.  They are
not sensitive to absolute neutrino masses.  We do know, from the atmospheric $|\delta m_{32}^2|$,
that at least one neutrino must have a mass $\gsim$ 0.05 eV.  But the best laboratory bound,
from tritium $\beta$ decay experiments, would allow neutrino masses 50 times greater.  That
is, the three light neutrinos might be nearly equal in mass, split by the tiny mass differences
indicated by solar and atmospheric neutrino oscillations.
\item Matter effects (from passage through the Earth) have not been seen in atmospheric
neutrino experiments.  This leaves open two possible orderings (normal, inverted) of the mass eigenstates, as
indicated above and as
illustrated in Fig.~\ref{fig:hierarchy}.  Atmospheric neutrino detectors that can determine the sign 
of the produced lepton are of potential interest in resolving the hierarchy question.
\item Very little is known about the three CP-violating phases that appear in the PMNS matrix. 
The Dirac phase could be determined
by looking for differences between certain conjugate oscillation channels, such as
$P(\nu_e \rightarrow \nu_\mu)$ and $P(\bar{\nu}_\mu \rightarrow \bar{\nu}_e)$, one of the goals
of future long-baseline neutrino experiments.  Finding
such a difference could be important to theories that attribute the excess of
matter over antimatter in our universe to leptonic CP violation (though the precise connection
between low-energy CP violation and the high-energy mechanism responsible for
baryogenesis may be difficult to define).
\item The neutrino, lacking an electric or any other charge that must flip sign under particle-antiparticle
conjugation, is unique among Standard-Model particles in that it may be its own antiparticle.  
So far no measurement has been made that can distinguish this possibility (a Majorana
neutrino) from the case where the $\nu$ and $\bar{\nu}$ are distinct (a Dirac neutrino).  
Next-generation neutrinoless double $\beta$ decay experiments
\[ (N,Z) \rightarrow (N-2,Z+2) + 2 e^- \]
could settle this issue, however.   This process requires lepton number violation and 
Majorana masses.  The two remaining CP-violating phases are Majorana phases
that can affect neutrinoless double $\beta$ decay rates.
\item No compelling argument has been given to account for the large mixing angles deduced
from atmospheric and solar neutrino oscillations.  These angles differ markedly from their
measured counterparts among the quarks.
\end{itemize}
While some of these questions may be answered in terrestrial experiments, neutrino
astrophysics will continue to offer unique environments for probing fundamental neutrino
properties.  Several examples are given in the chapter on neutrino cooling.

\section{Supernovae Neutrinos and Nucleosynthesis}
The bursts associated with a core collapse supernova are among
the most interesting sources of neutrinos in astrophysics \cite{models}.  A massive
star, in excess of 10 solar masses, begins its lifetime burning
the hydrogen in its core under the conditions of hydrostatic
equilibrium.  When the hydrogen is exhausted, the core contracts
until the density and temperature are reached where 3$\alpha \rightarrow
^{12}$C can take place.  The helium is then burned to exhaustion.
This pattern (fuel exhaustion, contraction, heating, and ignition of the 
ashes of the previous burning cycle) repeats several times,
leading finally to the explosive burning of Si to Fe.
For a heavy star, the evolution is rapid due to the amount of energy
the star must produce to support itself against its own gravity.
A 25-solar-mass star would go through
the set of burning cycles in about 7 My, with the final explosive Si 
burning stage taking a few days.  The result is an 
``onion skin" structure of the precollapse star 
in which the star's history can be read by looking at the 
surface inward: there are concentric shells dominated by H, ${}^4$He,
${}^{12}$C, ${}^{16}$O and ${}^{20}$Ne, ${}^{28}$Si, and ${}^{56}$Fe
at the center. 

\subsection{The Explosion Mechanism and Neutrino Burst}
The source of energy for this evolution is nuclear binding energy.
A plot of the nuclear binding energy $\delta$ as a function of nuclear
mass shows that the minimum is achieved at Fe.  In a scale
where the ${}^{12}$C mass is picked as zero:
\begin{center}
${}^{12}$C~~~~~$\delta$/nucleon = 0.000 MeV \\
${}^{16}$O~~~~~$\delta$/nucleon = -0.296 MeV \\
${}^{28}$Si~~~~$\delta$/nucleon = -0.768 MeV \\
${}^{40}$Ca~~~~$\delta$/nucleon = -0.871 MeV \\
${}^{56}$Fe~~~~$\delta$/nucleon = -1.082 MeV \\
${}^{72}$Ge~~~~$\delta$/nucleon = -1.008 MeV \\
${}^{98}$Mo~~~~$\delta$/nucleon = -0.899 Mev
\end{center}
Once the Si burns to produce Fe, there is no further source
of nuclear energy adequate to support the star.  So as the last
remnants of nuclear burning take place, the core is largely
supported by degeneracy pressure, with the energy generation rate
in the core being less than the stellar luminosity.  The core
density is about 2 $\times 10^9$ g/cc and the temperature is
kT $\sim$ 0.5 MeV. 

Thus the collapse that begins with the end of Si burning is
not halted by a new burning stage, but continues.  As gravity
does work on the matter, the collapse leads to a rapid heating
and compression of the matter.  Sufficient heating of the Fe can release $\alpha$s and a few
nucleons, which are bound by $\sim$ 8 MeV.  At the same time, the electron chemical potential is
increasing.  This makes electron capture on nuclei and any free
protons favorable,
\[ e^- + p \rightarrow \nu_e + n. \]
As the chemical equilibrium condition is
\[ \mu_e + \mu_p = \mu_n + \langle E_\nu \rangle, \]
the increase in
the electron Fermi surface with density will lead to
increased neutronization of the matter, as long as neutrinos
freely escape the star.  These escaping neutrinos carry
off energy and lepton number.  Both the electron capture and
the nuclear excitation and disassociation take energy out of the electron gas,
which is the star's only source of support.  Consequently
the collapse is very rapid, with numerical simulations finding that 
the star's iron core ($\sim$ 1.2-1.5 solar mases) collapses
at about 0.6 of the free-fall velocity.

While the $\nu_e$s readily escape in the early stages of infall,
conditions change once the 
density reaches $\sim$ 10$^{12}$g/cm$^3$. 
At this point neutrino scattering off the matter through
both charged current and coherent neutral current processes
begins to alter the transport.  The
neutral current neutrino scattering off nuclei is particularly
important, as the scattering amplitude is proportional to the total nuclear
weak charge, which is approximately the 
neutron number.  Elastic scattering transfers very little energy because
the mass of the nucleus is so much greater than the
typical energy of the neutrinos.  But momentum is exchanged. 
Because of repeated scattering the neutrino ``random walks" out of the star.  When the
neutrino mean free path becomes sufficiently short, the time
required for the neutrino to diffuse out of the high-density core
begins to exceed the time required to complete
the collapse.  Above densities of about
10$^{12}$ g/cm$^3$, or $\sim$ 1\% of nuclear density, such neutrino trapping occurs.
Consequently, once this critical density is exceeded, 
the energy released by further gravitational
collapse is trapped within the star until after core bounce.  
Similarly, the star no longer can lose lepton number due to neutrino emission.

For a neutron star of 1.4 solar masses and a radius of
10 km, an estimate of its binding energy is
\begin{equation}
 {G M^2 \over 2R} \sim 2.5 \times 10^{53} \mathrm{ergs}. 
\end{equation}
Thus this is roughly the trapped energy that will later be radiated in neutrinos,
after core bounce, as the proto-neutron star formed in the collapse cools. 

The collapse produces a shock wave that is critical to
subsequent ejection of the star's mantle.
The velocity of sound in matter rises with increasing density.
Late in the collapse the sound velocity in the inner portion of the iron core, 
with $M_{HC} \sim 0.6-0.9$ solar masses,
exceeds the infall velocity.  Any pressure
variations that may develop during infall can 
even out before the collapse is completed.  Consequently this
portion of the iron core collapses as a unit, retaining its density
profile.  

The collapse of the core continues until nuclear
densities are reached.  As nuclear matter is rather incompressible ($\sim$ 200 MeV/f$^3$),
the nuclear equation of state is effective in halting the collapse:
maximum densities of 3-4 times nuclear are reached, e.g.,
perhaps $6 \times 10^{14}$ g/cm$^3$.  The innermost shell of matter
reaches this supernuclear density first, rebounds, sending a 
pressure wave out through the inner core.  This wave
travels faster than the infalling matter, as the inner iron 
core is characterized by a sound speed in excess of the infall
speed.  Subsequent shells follow.  The resulting pressure
waves collect at the edge of the inner iron core -- the radius at which
the infall velocity and sound speed are equal.  
As this point reaches nuclear density and comes to
rest, a shock wave breaks out and begins its traversal of the 
outer core. 

Initially the shock wave may carry an order of magnitude more energy
than is needed to eject the mantle of the star (less than 10$^{51}$
ergs).  But as the shock wave travels through the outer iron core,
it heats and melts the iron that crosses the shock front, at a 
loss of $\sim$ 8 MeV/nucleon.  Additional energy is lost by neutrino
emission, which increases after the melting.   These losses are comparable to 
the initial energy carried by the shock wave.  Most simplified
(e.g., one dimensional) numerical models fail to produce a successful ``prompt"
hydrodynamic explosion, for this reason.   The shock stalls near the
edge of the iron core, instead of propagating into the mantle.

Most of the theoretical attention in the past decade has focused on the role
of neutrinos in reviving this shock wave, a process that becomes more 
effective in multi-dimensional models that account for convection.  In this delayed mechanism,
the shock wave stalls at a radius of 200-300 km, some tens of milliseconds
after core bounce.  But neutrinos diffusing out of the proto-neutron star
react frequently in the nucleon gas left in the wake of the shock wave,
depositing significant energy.  Over $\sim$ 0.5 seconds the increasing pressure
due to neutrino heating of this nucleon gas helps
push the shock outward.  This description is over-simplified -- a variety of
contributing effects are emerging from numerical simulations -- but there is
wide agreement that energy deposition by neutrinos is an essential ingredient
for successful explosions.

Regardless of explosion details, neutrinos dominate supernova energetics.
The kinetic energy of the explosion and
supernova's optical display account for less than 1\% of the available energy.
The remaining 99\% 
of the 3 $\times 10^{53}$ ergs released in the collapse is 
radiated in neutrinos of all flavors.  The timescale governing the leakage 
of trapped neutrinos out of the proto-neutron star
is about three seconds. 
The energy is roughly equipartitioned
among the flavors (a consequence of reactions
among trapped neutrinos that equilibrate flavor).  The detailed
decoupling of the emitted neutrinos from the matter -- which occurs at a
density of about $10^{11}-10^{12}$ g/cm$^3$ --does depend on flavor.
This leads to differences in neutrino temperatures, with electron
neutrinos being somewhat cooler ($T \sim$ 3.5 MeV) than the 
heavy-flavor neutrinos ($T \sim$ 6 MeV).
The radius for neutrino-matter decoupling defines a ``neutrinosphere" deep within the star, 
analogous to the familiar photosphere for optical emissions.

\begin{figure}
\begin{center}
\includegraphics[width=12cm]{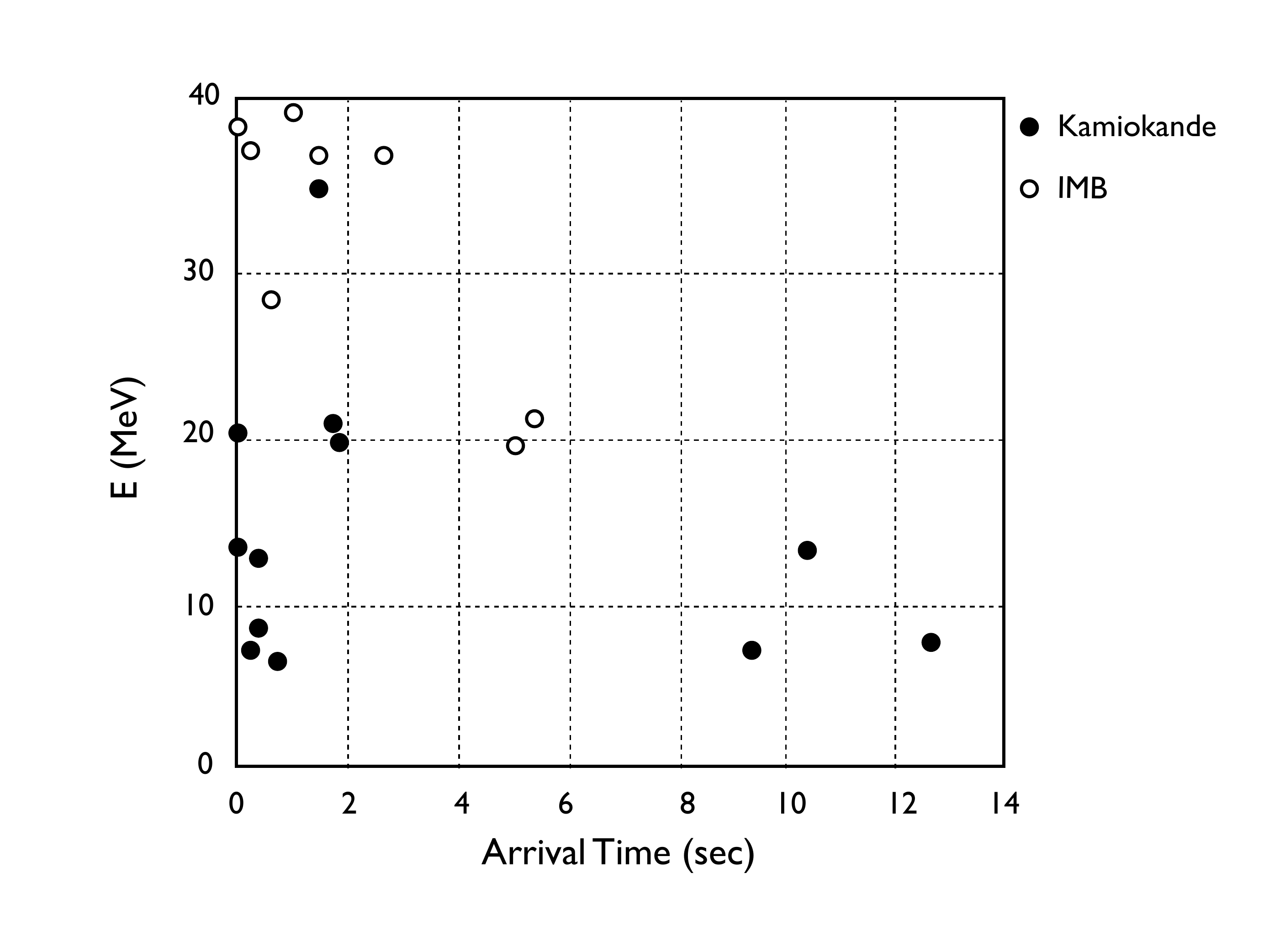}
\end{center}
\caption{ The timing and energy of neutrino events from SN1987A, as
observed in the Kamiokande and IMB detectors \cite{snnus}.}
\label{fig:SN1987Anew}
\end{figure}

The burst of neutrinos produced in a galactic core-collapse supernova
is detectable with instruments like Super-Kamiokande and SNO.   On February 23,
1987, a neutrino burst from a supernova in the Large Magellanic Cloud was
observed in the proton-decay detectors Kamiokande and IMB \cite{snnus}.   The optical
counterpart reached an apparent magnitude of about 3, and could be
observed easily in the night sky with the naked eye.  This supernova originated
160,000 light years from Earth.  Approximately 20 events were seen in the 
Kamiokande and IBM detectors, spread over approximately 10 seconds.  
Within the limited statistics possible with these first-generation detectors, the
number of events and the burst duration were consistent with standard 
estimates of the energy release and cooling time of a supernova.  The
neutrino data from the two detectors are shown in Fig.~\ref{fig:SN1987Anew}.

\begin{figure}
\begin{center}
\includegraphics[width=12cm]{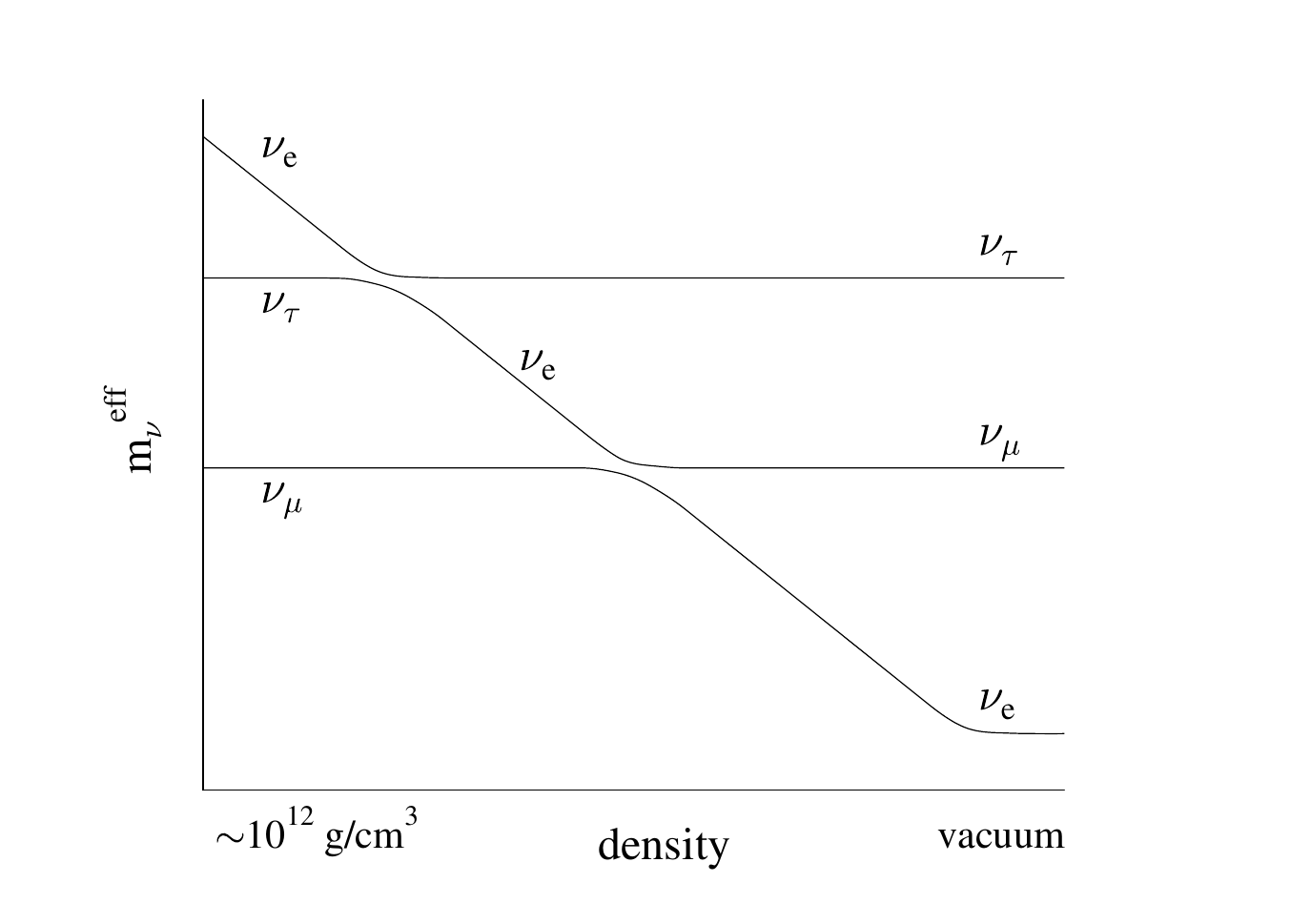}
\end{center}
\caption{A schematic illustration of the two level crossings that the $\nu_e$ would
experience, given the higher densities available in the mantle of a massive
star undergoing core collapse.}
\label{fig:threelevel}
\end{figure}

Temperature differences between neutrino flavors are interesting
because of oscillations and nucleosynthesis.  The discussion of matter
effects in the solar neutrino problem was limited to two flavors.  But the
higher densities found in core-collapse supernovae make all
three flavors relevant.  The three-flavor MSW
level-crossing diagram is shown in Fig.~\ref{fig:threelevel}.  One sees, in
addition to the neutrino ``level crossing"  $\nu_\mu \leftrightarrow \nu_e$
important for solar neutrinos, a second crossing of the $\nu_e$ with the
$\nu_\tau$.  The higher density characterizing this second crossing,
$\sim 10^4$ g/cm$^3$, is determined by atmospheric mass difference $|\delta m_{32}^2|$
and by the typical energy of supernova neutrinos, $\sim 10-20$ MeV.
This density is beyond that available in the Sun ($\lsim 10^2$ g/cm$^3$),
but far less than that of a supernova's neutrinosphere.  Consequently
this second level crossing alters neutrino flavor only after the supernova
neutrinos are free-streaming out of the star, with well defined spectra
that are approximately thermal.   This level crossing can exchange the
flavor labels on the $\nu_e$ and $\nu_\tau$ spectra, so that the $\nu_e$s
become hotter than the heavy-flavor neutrinos.  One would expect such 
an inversion to be apparent in terrestrial supernova neutrino detectors.

In fact this description oversimplifies the neutrino physics of supernovae.
The enormous neutrino densities encountered in a supernova lead to a new 
aspect of the MSW effect -- oscillations altered not by neutrino-electron
scattering, but by neutrino-neutrino scattering \cite{fuller}.   While the precise consequences of
this neutrino-neutrino MSW potential are still being explored -- the problem is 
both nonlinear and dependent on the angles and flavors of interacting neutrinos -- the effects reach
much deeper into the star and alter the flavor physics in distinctive ways.  
Consequently, this novel flavor physics could play a role in the dynamics of the explosion.
Supernovae likely provide the only environment in nature where
neutrino-neutrino interactions dominate the MSW potential.

\subsection{Supernova Neutrino Physics}
This novel neutrino-neutrino MSW potential is one of many reasons core-collapse supernovae
play an important role in neutrino astrophysics.  Others include:
\begin{itemize} 
\item As $\sim$ 99\% of the collapse energy is radiated in neutrinos, one can in principle
deduce the the binding energy of the neutron star from neutrino flux measurements,
provided other parameters (such as the distance to the supernova) are sufficiently well known.
The recent observation of a two-solar-mass neutron star suggests that the nuclear 
equation of state is stiff at high density, disfavoring very compact neutron stars.
\item Neutrinos from galactic supernovae will not be obscured by intervening matter or
dust, unlike optical signals.  Thus supernova neutrino bursts should, over the next few
hundred years, provide our most reliable measure of the contemporary rate of
galactic core collapse.
\item There exists a so-far undetected diffuse background of supernova neutrinos, produced
by all past supernovae occurring in the universe.  Future detectors that may approach
the megaton scale should be able to see a few events from this source.  Detection of
these neutrinos would place a important constraint on the inventory of massive stars
undergoing core collapse, from the first epoch of star formation until now.
\item Supernovae are one of the most important engines for nucleosynthesis, controlling
much of the chemical enrichment of the galaxy.  As described in the next section,
neutrinos are directly and indirectly involved in this synthesis.
\item While most of the energy released in core collapse is radiated as neutrinos over the
first several seconds, neutrino emission at a lower level continues as the proto-neutron
star cools and radiates away its lepton number.  It is quite possible that phase changes
in the dense nuclear matter could occur several tens of seconds after core bounce,
altering the late-time neutrino ``light curve" in a characteristic way.   
While there are considerable uncertainties in
estimates, neutrino processes may continue to dominate neutron star cooling for
$\sim 10^5$ years.
\item  The neutrino burst could include other sharp features in time, marking interesting
astrophysics.  The melting of iron to nucleons with the passage of the shock wave through
the outer iron core is predicted to produce a spike in the neutrino luminosity, lasting for
a few milliseconds.  Continued accretion onto the neutron star surface could produce a
collapse to a black hole, and consequently a sudden termination in neutrino emission.
\item  Supernova cooling times place constraints on new physics associated with particles
that also couple weakly to matter.  For example, a light scalar called the axion could,
in principle, compete with neutrinos in cooling a supernovae.  The requirement that axion
emission not shorting the cooling time too much, which would be in conflict with SN1987A
data, constrains the mass and coupling of this hypothesized particle.
\end{itemize}

\section{Neutrinos and Nucleosynthesis}
Neutrinos and nucleosynthesis are both associated with explosive environments 
found in astrophysics.  This section discusses three examples, the Big Bang, the
neutrino process, and the r-process.

\subsection{Cosmological Neutrinos, BBN, and Large-scale Structure}
One of the classic problems in nucleosynthesis and cosmology is accounting for
the mass fraction of primordial helium of 25\%, as well as the
abundances of D, ${}^3$He,  and ${}^7$Li.  The ${}^4$He mass fraction
shows that nuclei were synthesized from the
early-universe nucleon soup at a time when the n/p ratio was $\sim$ 1/7.   This requires
the proper coordination of two ``clocks," one being the weak interaction rate for decay of
neutrons to protons, and the other the Hubble rate governing the expansion of the
universe,
\[ H(t) \equiv {1 \over R(t)} {d R(t) \over dt} = \sqrt{ 8 \pi G \rho(t) \over 3}, \]
where $G$ is the gravitational constant.
This clock depends on the energy density $\rho(t)$ which, at this epoch, is 
dominated by relativistic particles, including the neutrinos.   Big Bang nucleosynthesis (BBN)
depends on one adjustable parameter, the baryon density, which is usually given
in terms of the ratio of baryons to photons $\eta$.   Perhaps the first quantitative
result in cosmology was the BBN determination of $\eta \sim (5.9 \pm 0.8) \times 10^{-10}$ \cite{Steigman},
from which one can deduce the baryon density $\rho_b \sim 4.2 \times 10^{-31}$ g/cm$^3
\sim 0.044 \rho_\mathrm{crit}$, where $\rho_\mathrm{crit}$ is the closure density.
The comparison of $\rho_b$ with various astrophysical measurements of the
amount of gravitating matter in the universe shows that approximate 85\% of
that matter is dark, nonbaryonic, and consequently outside the Standard Model.

An independent and more precise determination of $\eta$ has been made
from the pattern of acoustic peaks seen in mappings of the cosmic microwave
background.  The seven-year WMAP result is  $(6.19 \pm 0.15) \times 10^{-10}$ \cite{WMAP7}.
The two determinations, one 
based on the nucleosynthesis that took place three minutes after the Big Bang and the
second connected with the pattern of large-scale structure at the time
the first atoms formed 380,000 years later,
are in remarkable agreement.

The primordial abundance of ${}^4$He is
relatively insensitive to $\eta$, in contrast to other BBN species like D.   Consequently this 
abundance is a rather good test of the
number of neutrino flavors contributing to early-universe expansion.  BBN analyses over 
a number of years have generally favored $N_\nu \lsim 3$, though calculations using $N_\nu =3$ 
yield the observed abundances of 
both D and $^4$He at 68\% c.l. \cite{Steigman}.  However, recent reanalyses of the $^4$He
abundance \cite{Iz} have led to an upward revision, so that $N_\nu \sim 3.74^{+0.86}_{-0.76}$.
The seven-year  WMAP analysis of the CMB constrains a similar quantity,
the effective number of relativistic species, yielding $N_\mathrm{eff} \sim 4.34 \pm 0.87$,
where $\sim 3$ is expected \cite{WMAP7}.  This modest discrepancy 
as well as the disagreement between the
BBN primordial $^7$Li prediction and Li abundances deduced from old, metal-poor stars
have stimulated recent interest in nonstandard BBN scenarios \cite{Field,Pospelov}.

Because neutrinos are massive and relativistic in the early universe, they are an
interesting component of dark matter.  The lighter the scale of neutrino mass, the longer
they remain relativistic, the further they stream, and the more effective they are in
suppressing the growth of large-scale structure.   The evolution of neutrino mass effects
on structure growth is distinctive in both red shift Z and
spatial scale, altering the distribution of baryons and cold dark matter at the $\sim$ 1\% level when their
contribution to the critical density is just $\Omega_\nu \sim 0.1\%$.   The seven-year WMAP
analysis yielded \cite{WMAP7}
\[ \sum_i m_\nu(i) \lsim 0.58~ \mathrm{eV}, \]
though tighter bounds have been claimed in other, combined analyses.
A variety of cosmological surveys planned for the next decade
have the anticipated statistical power to test neutrino mass effects with an order of
magnitude or more increase in sensitivity \cite{Abazajian}.   As the atmospheric neutrino mass difference 
implies a lower bound on neutrino mass of $\sim$ 50 meV, the absolute scale of neutrino
mass could be determined cosmologically.  A cosmological bound significantly below
100 meV would also imply that the hierarchy is normal.

These arguments assume that cosmological measurements of the energy density in
relativistic species constrain Standard-Model neutrinos.  There are a number of possible exceptions
to this assumption that are of significant current interest, such as the consequences of
a net lepton number asymmetry in the universe, or the presence of a sterile neutrino
(a neutrino lacking Standard-Model couplings) that might mix with active species. 

\subsection{The Neutrino Process}
One of the more amusing roles for neutrinos in nucleosynthesis is found in
the neutrino process, the direct synthesis of new elements through neutrino reactions.
Core-collapse supernovae provide the enormous neutrino fluences
necessary for such synthesis to be significant.  They also eject newly 
synthesized material into the interstellar medium, where it can be incorporated
into a new generation of stars.

Among the elements that might be made primarily or partially in the $\nu$-process,
the synthesis of ${}^{19}$F is one of the more interesting examples \cite{woosley}.
The only stable isotope of fluorine,
${}^{19}$F has an abundance
\[ {^{19}\mathrm{F} \over ^{20}\mathrm{Ne}} \sim {1 \over 3100}. \]
Ne is one of the hydrostatic burning products in massive stars, produced in
great abundance and ejected in core-collapse supernovae.  Thus a mechanism
that converts $\sim$ 0.03\% of the ${}^{20}$Ne in the star's mantle into ${}^{19}$F could
account for the entire observed abundance of the latter.

The Ne zone in a supernova progenitor star
is typically located at a radius of $\sim$ 20,000 km.  A simple calculation
that combines the neutrino fluence through the Ne zone
with the cross section for inelastic neutrino scattering off ${}^{20}$Ne shows that
approximately 0.3\% of the ${}^{20}$Ne nuclei would interact with the
neutrinos produced in the core collapse.  Almost all of these
reactions result in the production of ${}^{19}$F, e.g., 
\begin{eqnarray}
 {}^{20}\mathrm{Ne}(\nu,\nu')^{20}\mathrm{Ne}^* &\rightarrow& {}^{19}\mathrm{Ne} + n 
\rightarrow ^{19}\mathrm{F} + e^+ + \nu_e + n \nonumber \\
  {}^{20}\mathrm{Ne}(\nu,\nu')^{20}\mathrm{Ne}^* &\rightarrow& {}^{19}\mathrm{F}
+ p, \nonumber
\end{eqnarray}
with the first reaction occurring half as frequently as the 
second.  Thus one would expect the abundance
ratio to be ${}^{19}\mathrm{F}/ {}^{20}\mathrm{Ne} \sim 1/ 300$, corresponding to an order
of magnitude more ${}^{19}$F than found in nature.  

This example shows that stars are rather complicated factories for nucleosynthesis.
Implicit in the reactions above are mechanisms that also destroy ${}^{19}$F.  
For example, about 70\% of the neutrons coproduced with ${}^{19}$F in the first reaction
immediately recapture on ${}^{19}$F, destroying the product of interest.  
Similarly, many of the coproduced protons destroy ${}^{19}$F via ${}^{19}$F(p,$\alpha)^{16}$O --
unless the star is rich in ${}^{23}$Na, which readily consumes protons via
${}^{23}$Na(p,$\alpha)^{20}$Ne.  Finally, some of the ${}^{19}$F produced in the
neon shell is destroyed when the shock wave passes through that zone:
the shock wave can heat the inner portion of the Ne zone above $1.7 \times 10^9$K, the
temperature at which ${}^{19}$F can be destroyed by ${}^{19}$F($\gamma,\alpha)^{15}$N.

If all of this physics is treated carefully in a nuclear network code, one finds that the
desired ${}^{19}$F/${}^{20}$Ne $\sim$ 1/3100 is achieved for a heavy-flavor neutrino
temperature of about 6 MeV.  This is quite consistent with the temperatures that come
from supernova models.

The neutrino process produces interesting abundances of several relatively
rare, odd-A nuclei including ${}^7$Li, ${}^{11}$B, 
${}^{138}$La, ${}^{180}$Ta, and ${}^{15}$N.  Charged-current neutrino reactions
on free protons can produce neutrons that, through $(n,p)$ and $(n,\gamma)$
reactions, lead to the nucleosynthesis of the so-called ``p-process" nuclei from
A=92 to 126.  The production of such nuclei has been a long-standing puzzle in nuclear astrophysics.

\subsection{The r-process}
Beyond the iron peak nuclear Coulomb barriers become so high
that charged particle reactions become ineffective, leaving
neutron capture as the mechanism responsible for producing
the heaviest nuclei.
If the neutron abundance is modest,
this capture occurs in such a way that each newly synthesized
nucleus has the opportunity to $\beta$ decay, if it is energetically
favorable to do so.  Thus weak equilibrium is maintained among
the nuclei, so that synthesis is along the nuclear valley of stability.
This is called the s- or slow-process.  However a
plot of the s-process in the (N,Z) plane reveals that this
path misses many stable, neutron-rich nuclei.  This suggests that another mechanism is also at
work.  Furthermore, the abundance peaks found in nature 
near masses A $\sim$ 130 and A $\sim$ 190, which mark the closed
neutron shells where neutron capture rates and $\beta$ decay
rates are slower, each split into two subpeaks.  One set of subpeaks
corresponds to the closed-neutron-shell numbers N $\sim$ 82
and N $\sim$ 126, and is clearly associated with the s-process.
The other set is shifted to smaller N $\sim$ 76 and $\sim$ 116,
respectively, suggestive of a much more explosive
neutron capture environment. 
  
This second process is the r- or rapid-process \cite{qian}, which requires a neutron fluence
so large that neutron capture is fast compared to $\beta$ decay.  In this case
nuclei rapidly absorb neutrons until they approach the neutron drip line.  That is,
equilibrium is maintained by $(n,\gamma) \leftrightarrow
(\gamma,n)$, not by weak interactions.  Consequently the nuclei participating in the r-process are very
different from ordinary nuclei -- very neutron rich nuclei that would decay immediately
in the low-temperature environment of Earth.   The rate of nucleosynthesis is
controlled by the rate of $\beta$ decay: a new neutron can be captured only
after $\beta$ decay, $n \rightarrow p + e^- +\bar{\nu}_e$,
opens up a hole in the neutron Fermi sea.  Consequently one expects abundance
peaks near the closed neutron shells at N $\sim$ 82 and 126, as $\beta$ decay 
is slow and mass will pile up at these ``waiting points."  By a series of rapid neutron captures 
and slower $\beta$ decays, synthesis can proceed all the way to the transuranics. 
Typical r-process conditions include neutron
densities $\rho(n) \sim 10^{18}-10^{22}$/cm$^3$, temperatures $\sim$
10$^9$ K, times $\sim$ 1 second, and ratios of free neutrons to heavy seed 
nuclei of $\gsim$ 100 (so that the transuranics can be synthesized from Fe).
  
The path of the r-process is typically displaced by just $\sim$ (2-3) MeV from
the neutron drip line (where no more bound neutron levels exist).
After the r-process neutron exposure ends,
the nuclei decay back to the valley of stability by $\beta$
decay.  This involves conversion of neutrons into protons,
shifting the r-process peaks from the parent-nucleus values of N $\sim$ 82 and 126
to lower values and explaining the double-peak structure of the 
r-process/s-process closed-shell abundance peaks.

\begin{figure}
\begin{center}
\includegraphics[width=14cm]{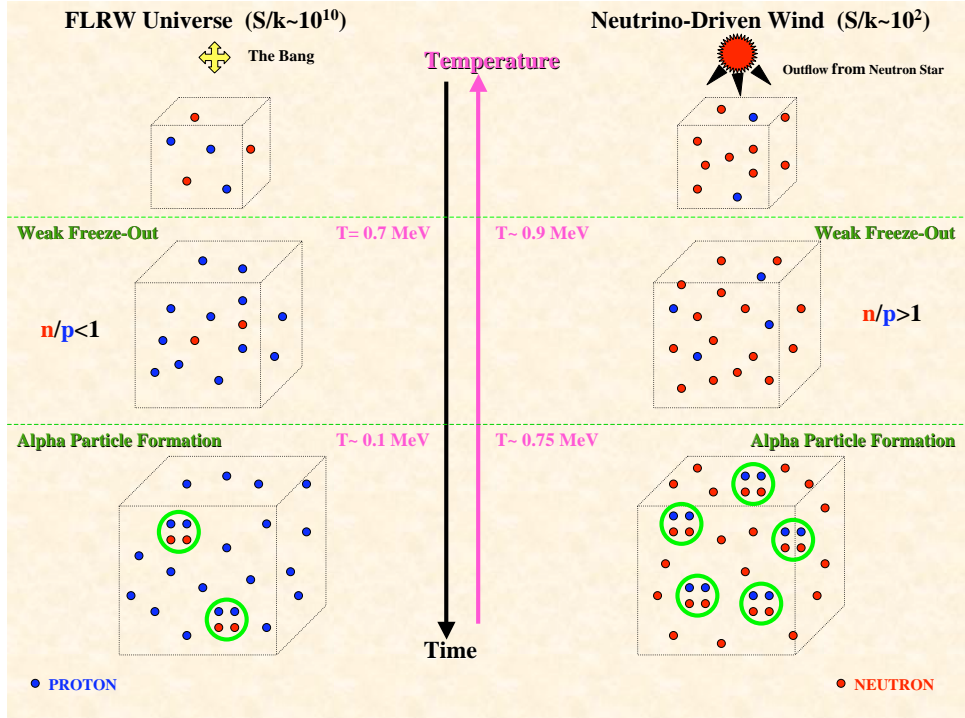}
\end{center}
\caption{ A comparison of two remarkably similar mechanisms for nucleosynthesis,
an expanding, cooling, high entropy, proton-rich nucleon gas (the Big Bang, left panel)
and an expanding, cooling, high entropy, neutron-rich nucleon gas (the supernova wind r-process,
right panel).  In the former, the synthesis
determinates at $^4$He and a few other light elements. In the latter, the
synthesis continues to the heavy transuranic elements.  Figure from G. Fuller.}
\label{fig:rprocess}
\end{figure}

One possible neutrino role in the r-process is in producing the required
explosive, neutron-rich environments.   One such site is the supernova neutrino-driven wind -- the
last ejecta blown off the proto-neutron star.
This material is hot, dominated by radiation, and
contains neutrons
and protons, often with an excess of neutrons.  As the nucleon gas expands
off the star and cools, it goes through
a freezeout to $\alpha$ particles, a step that essentially
locks up all the protons.
Then the $\alpha$s interact through reactions like 
\[ \alpha + \alpha +\alpha \rightarrow ^{12}C  \]
to start forming heavier nuclei.  The
$\alpha$ capture continues, eventually synthesizing intermediate-mass
``seed" nuclei.  Once these seed nuclei are produced, if the requisite number of
neutrons is available ($\sim$ 100 per seed nucleus), very heavy nuclei
can be made.   The scenario, as depicted in Fig.~\ref{fig:rprocess},
is quite similar to the Big Bang:  a hot nucleon gas (with an entropy
$S\sim$ 100 in Boltzmann units, compared to the BBN $S\sim$ 10$^{10}$)
expand and cools, condensing into nuclei.  But a detail -- the neutrino
wind has an excess of neutrons, while the Big Bang is proton-rich -- leads
to uranium in one case, and to termination of nucleosynthesis
at $^4$He (plus a few light elements) in the other.  The
neutrinos are crucial: they help keep the entropy of the nucleon
gas high, control the n/p ratio
of the gas through competing charge-current reactions, and generate the wind
that ejects the r-process products.

There are some very nice aspects of this site: the amount of
matter ejected is $\sim 10^{-6}$ solar masses,
a production per event that if integrated over the lifetime of the
galaxy gives the required total abundance of r-process metals,
assuming typical supernova rates.  There are also
some significant problems -- including great difficulties in maintaining
the necessary neutron/seed ratio of $\gsim$ 100 for realistic wind
entropies \cite{Roberts} -- that have encouraged investigations of other sites.
Possibilities include a so-called ``cold" neutrino-driven r-process
operating in the ${}^4$He mantles of early, metal-poor supernova progenitors \cite{Banerjee}
and neutron star mergers, which could be the dominant r-process
site once our galaxy has evolved to the point that these events become common \cite{mergers}.

\section{Neutrino Cooling and Red Giants}
Several neutrino cooling scenarios have already been discussed,
including cooling of the proto-neutron star produce in core collapse 
and cooling connected with the expansion of the early universe.
Red giant cooling provides an additional example of the use of 
astrophysical arguments to constrain
fundamental properties of neutrinos.

\subsection{Red Giants and Helium Ignition}
In a solar-like star, when the hydrogen in the central core has been
exhausted, an interesting evolution ensues:
\begin{itemize}
\item  With no further means of producing energy, the core
slowly contracts, thereby increasing in temperature as gravity
does work on the core.
\item Matter outside the core is still hydrogen rich, and
can generate energy through hydrogen burning.  Thus a hydrogen-burning
shell forms, generating the gas pressure supporting the outside layers of the
star.  As the ${}^4$He-rich core contracts, the matter outside the
core is also pulled deeper into the gravitational potential.
Furthermore, the H-burning shell continually adds more mass to the core.
This means the burning in the shell must intensify to generate
the additional gas pressure to fight gravity.  The shell also
thickens, as more hydrogen is above the
burning temperature.
\item The resulting increasing gas pressure causes the outer
envelope of the star to expand by a large factor, up to a 
factor of 50.  The increase in radius more than compensates for
the increased internal energy generation, so that a cooler
surface results.  The star reddens.  Stars of this type are
called red supergiants.
\item This evolution is relatively rapid, perhaps a few
hundred million years: the dense core requires large energy
production.  The helium core is supported
by its degeneracy pressure, and is characterized by densities
$\sim 10^6$ g/cm$^3$.  This stage ends when the
core reaches densities and temperatures that allow helium burning
through the reaction
\[ \alpha + \alpha + \alpha \rightarrow ^{12}C + \gamma . \]
As this reaction is quite temperature dependent,
the conditions for ignition are very sharply defined.
This has the consequence that the core mass at the helium flash point
is well determined.
\item  The onset of helium burning produces a new source of
support for the core.  The energy released elevates the temperature
and the core expands: He burning, not electron degeneracy, now 
supports the core.  The burning shell and envelope move
outward, higher in the gravitational potential.  Thus shell
hydrogen burning slows (the shell cools) because less gas pressure
is needed to satisfy hydrostatic equilibrium.  All of this
means the evolution of the star has now slowed: the red giant
moves along the ``horizontal branch," as interior temperatures
increase slowly, much as in the main sequence.
\end{itemize}

The 3$\alpha$ process involves some fascinating nuclear physics
that will not be recounted here:  the existence of certain nuclear resonances
was predicted based on the astrophysical requirements for this process.
The resulting He-burning rate exhibits a sharp temperature dependence
$\sim$ T$^{40}$ in the range relevant to red giant cores.
This dependence is the reason the He flash
is delicately dependent on conditions in the core.

\subsection{Neutrino Magnetic Moments and He Ignition}
Prior to the helium flash, the degenerate He core radiates
energy largely by neutrino pair emission.  The process is
the decay of a plasmon --- which one can think of as a photon
``dressed" by electron-hole excitations, thereby given the photon 
an effective mass of $\sim$ 10 keV.  The plasmon couples to
an electron particle-hole pair that
then decays via $Z_\mathrm{o} \rightarrow \nu \bar{\nu}$. 

If this cooling is somehow enhanced, the degenerate helium core 
would not ignite at the normal
time, but instead continue to grow.    When the core does finally
ignite, the larger core will alter
the star's subsequent evolution.

One possible mechanism for enhanced cooling is a neutrino
magnetic moment.  Then the plasmon could directly couple to
a neutrino pair.  The strength of this coupling would 
depend on the size of the magnetic moment.

A delay in the time of He ignition has several observable
consequences, including changing the ratio of red giant to
horizontal branch stars.  Thus, using the standard theory of
red giant evolution, investigators have attempted to determine
what size of magnetic moment would produce unacceptable 
changes in the astronomy.  The resulting limit \cite{raffelt} on diagonal or
transition neutrino magnetic moments,
\[ \mu_{ij} \lsim 3 \times 10^{-12} \mathrm{~electron~Bohr~magnetons} ,\]
is about an order of magnitude more stringent than the best limits
so far obtained from reactor neutrino experiments \cite{GEMMA}.

This example is just one of a number of such constraints that
can be extracted from similar stellar cooling arguments.
The arguments above, for example, can be repeated for 
neutrino electric dipole moments, or for
axion emission from red giants.   As noted previously, the
arguments can be extended to supernovae: anomalous
cooling processes that shorten the cooling time in a way that is
inconsistent with SN1987A observations are ruled out.  
For example, large Dirac neutrino masses are in conflict with SN1987A observations:
the mass term would allow neutrinos to scatter into sterile right-handed states,
which would then immediately escape, carrying off energy.

\section{High Energy Astrophysical Neutrinos}
\begin{figure}
\begin{center}
\includegraphics[width=12cm]{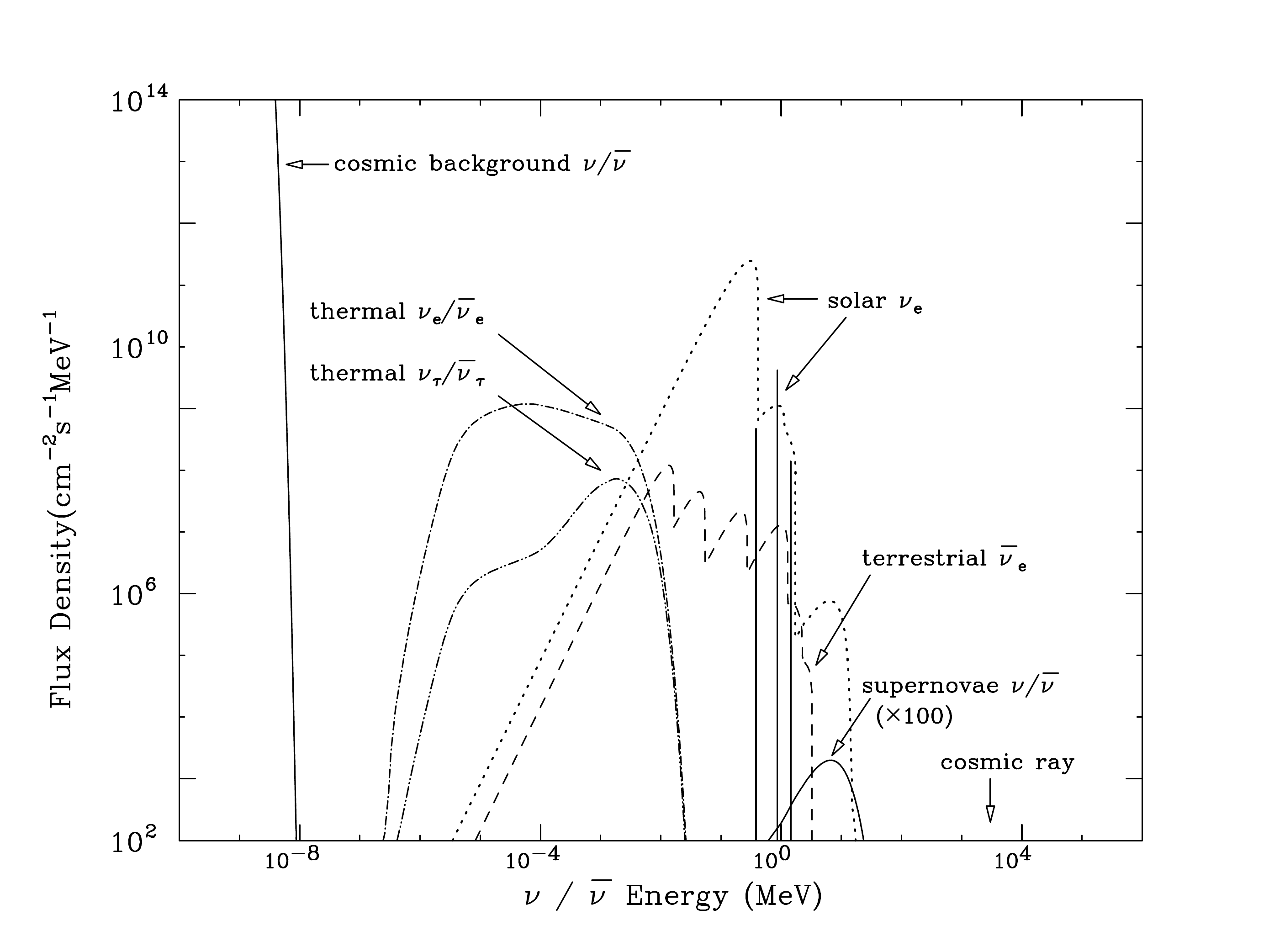}
\end{center}
\caption{The low-energy neutrino ``sky."  In addition to the principal sources
discussed in the text (Big-Bang, solar, supernova, and atmospheric neutrinos), the
natural neutrino background includes $\sim$ 1 keV thermal neutrinos of all flavors emitted
by the Sun as well as $\bar{\nu}_e$s generated by the Earth's natural radioactivity.
From \cite{lin}.}
\label{fig:lowEnus}
\end{figure}

\begin{figure}
\begin{center}
\includegraphics[width=14cm]{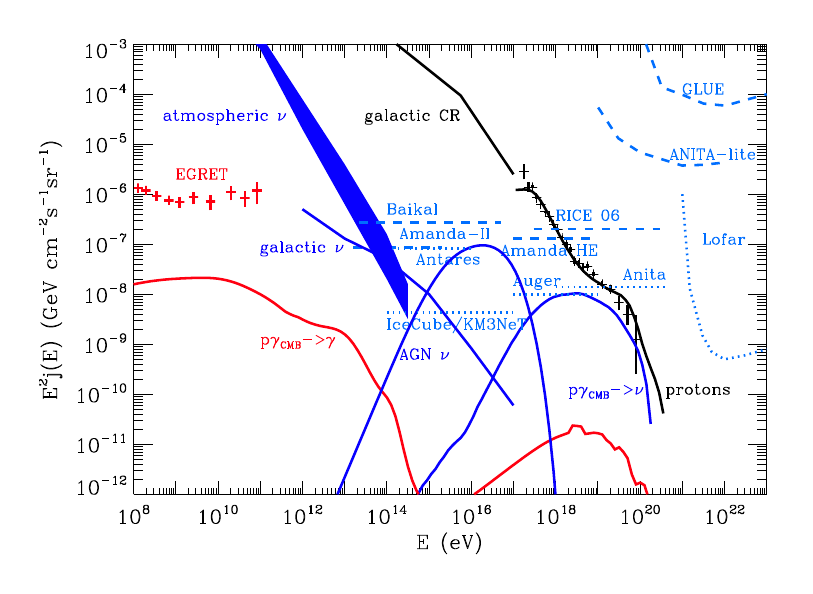}
\end{center}
\caption{A theoretical model of high-energy neutrino ``sky."  Shown are the expected
neutrino fluxes (blue area and lines), the primary cosmic ray fluxes (determined from data
and a model, black), and the secondary $\gamma$-ray fluxes expected from protons interacting
with the microwave background (red).  The neutrino flux is per flavor and includes
only relatively certain sources: atmospheric neutrinos, neutrinos resulting from cosmic-ray
interactions in our galaxy, and GZK neutrinos resulting from cosmic-ray interactions
with the microwave and infrared background.  The figure includes experimental data,
limits, and projected sensitivities to existing and planned telescopes.  Figure from G. Sigl \cite{sigl}.}
\label{fig:highEnusnew}
\end{figure}

Previous discussions focused on astrophysical neutrino sources
with energies ranging from the cosmic microwave temperature to $\lsim$ 10 GeV, 
a range including the bulk of atmospheric neutrinos.  These sources are displayed in Fig.~\ref{fig:lowEnus}
according to their contributions to the terrestrial flux density.  The figure includes low-energy sources,
such as the thermal solar neutrinos of all flavors, not explicitly discussed here
because of space limitations.  Beyond the figure's high-energy limits there exist
neutrino sources associated with
some of nature's most spectacular natural accelerators.  The program of
experiments to map out the high-energy neutrino spectrum
is just beginning.   Some guidance is provided by existing data on cosmic-ray protons, nuclei, and $\gamma$ rays, which constrain possible neutrino fluxes (see Fig.~\ref{fig:highEnusnew}). This concluding section discusses some of the suggested sources and current efforts 
to develop high-energy neutrino telescopes appropriate to these
sources.  The high-energy spectrum is one of the frontiers of neutrino astronomy.

\subsection{Neutrino Production by Cosmic Rays}
\begin{figure}
\begin{center}
\includegraphics[width=12cm]{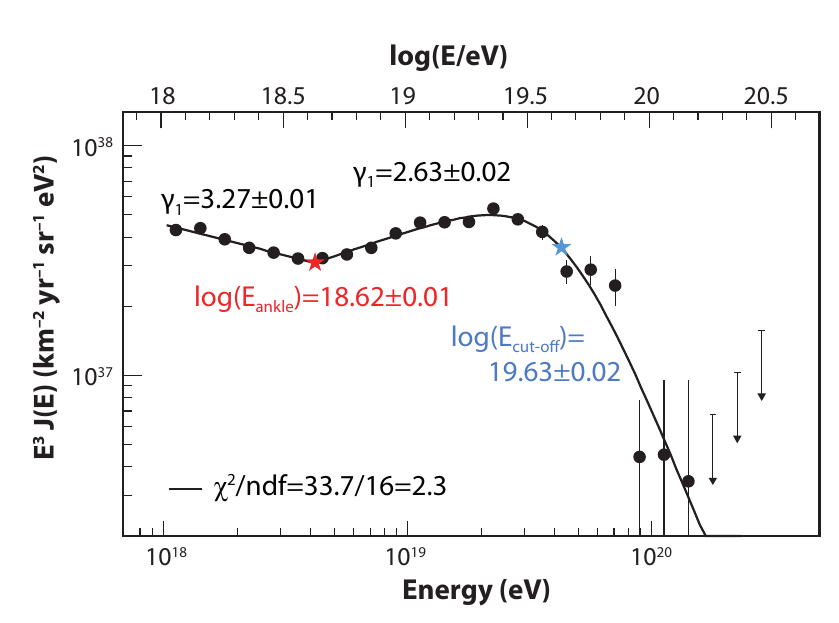}
\end{center}
\caption{Pierre Auger Observatory cosmic ray energy spectrum, fitted with two power laws
in the lower-energy ``ankle" region and with a smooth curve in the higher energy GZK region, where
a cut-off energy is identified.  In addition to the statistical uncertainties shown, there is a
systematic uncertainty in the energy scale of 22\%.  From \cite{PA11}, with permission 
of the Pierre Auger Observatory.}
\label{fig:pierre}
\end{figure}

The ultra-high-energy cosmic ray (UHECR) spectrum -- presumably protons and nuclei -- is known to
vary smoothly up to an energy $E \sim 4 \times 10^{19}$ eV.  The spectrum just below
this point is characterized by a spectral index $\alpha \sim -2.7$: $\phi_\nu(E) \sim
E^{\alpha}$.  Higher energy events
are seen, but the flux  drops off steeply beyond this point.  This is consistent with the
prediction of Greisen, Zatsepin, and Kuz'min \cite{GZK}: Above this cutoff UHECRs can loose energy by scattering off microwave photons, producing pions.  This sharply reduces 
the mean-free path of such UHECRs.  The declining flux reflects the reduced number of sources
within a mean-free path of the Earth.  This behavior has been mapped by the Pierre Auger
cosmic ray observatory.  The data shown in Fig. \ref{fig:pierre} 
can be reproduced with a smooth fitting function having a cutoff in the spectrum above 
$\sim (4.3 \pm 0.2) \times$ 10$^{19}$ eV \cite{PA11}.  Above the cutoff the primary signature
of UHECR protons and nuclei would be the neutrinos they produce when interacting with the CMB.

From models of the cosmic-ray spectrum of protons and nuclei, an
estimate can be made of the flux of UHE neutrinos
associated with the decays of pions and other secondaries produced in the GZK
scattering.  Uncertainties in this estimate include the flux, spectrum, and composition
of the UHECRs, the behavior of the spectrum beyond the GZK cutoff (as
we are blind to these UHECRs), and the spectrum's cosmological evolution.  One bound was obtained by Bahcall and Waxman \cite{waxman}; another is shown
in Fig.~\ref{fig:highEnusnew}.

These uncertainties are connected to some very interesting astrophysical issues:
the maximum energies that can be reached in
astrophysical accelerators; the UHECR uniformity over time (or equivalently redshift); and the role
other background photon sources, such as the infrared and optical backgrounds, in
producing UHE neutrinos from UHECRs.  A
more extended discussion can be found in \cite{APS}.

As in the case of low-energy sources such as solar neutrinos, the detection of very high
energy astrophysical neutrinos would open up new opportunities in both astrophysics
and particle physics.  Because the GZK cutoff limits the range of the UHECRs, neutrinos
provide the only direct probe of nature's most energetic astrophysical accelerators.  
The cutoff mechanism itself -- very high energy cosmic ray interactions with CMB photons
that can photo-produce pions and other secondaries -- is a source of very energetic neutrinos.
Because neutrinos travel in straight lines through magnetic fields, they point back to
their sources, allowing astronomers to correlate those sources with their optical
counterparts -- the accretion disks surrounding supermassive black holes, quasars, $\gamma$-ray bursts, etc.  The interactions of such energetic neutrinos with matter are untested experimentally,
as terrestrial accelerators have reach only the TeV scale.

\subsection{Cosmic Ray Studies, Point Sources, and Neutrino Telescopes}
The possibility of point sources is generally considered
 the astrophysical ``driver" for developing instruments to measure the highest
energy neutrinos.   There are intensely energetic sources in the sky, including 
active galactic nuclei (AGNs),
supernovae and associated phenomena like $\gamma$-ray bursts, 
and compact objects such as black holes and neutron
stars.    The magnetic fields, shock waves, gravitational fields, and energy densities associated
with such objects are
beyond those that can be produced in the laboratory.  

The measurements that have been made by the Pierre Auger Observatory, which began
operations in 2004 and reached its full scope in 2008, provide an important baseline
for high-energy neutrino studies.  Its observation of a high-energy cutoff consistent with
GZK predictions is one important step.  The high-energy behavior of the flux is consistent with either
a primary cosmic-ray spectrum dominated by protons or one dominated by heavier
nuclei, so that additional information on the character of air-shower observables is
needed to constrain composition (which influences GZK neutrino production).  Changes
in air-shower chartacteristics
above $E \gsim 5 \times 10^{18}$ eV indicate an increasingly
heavier composition.  The UHECR production of neutrinos can be significantly lower
if the UHECR primary spectrum is dominated by CNO or Fe nuclei \cite{Olinto}.

\begin{figure}
\begin{center}
\includegraphics[width=12cm]{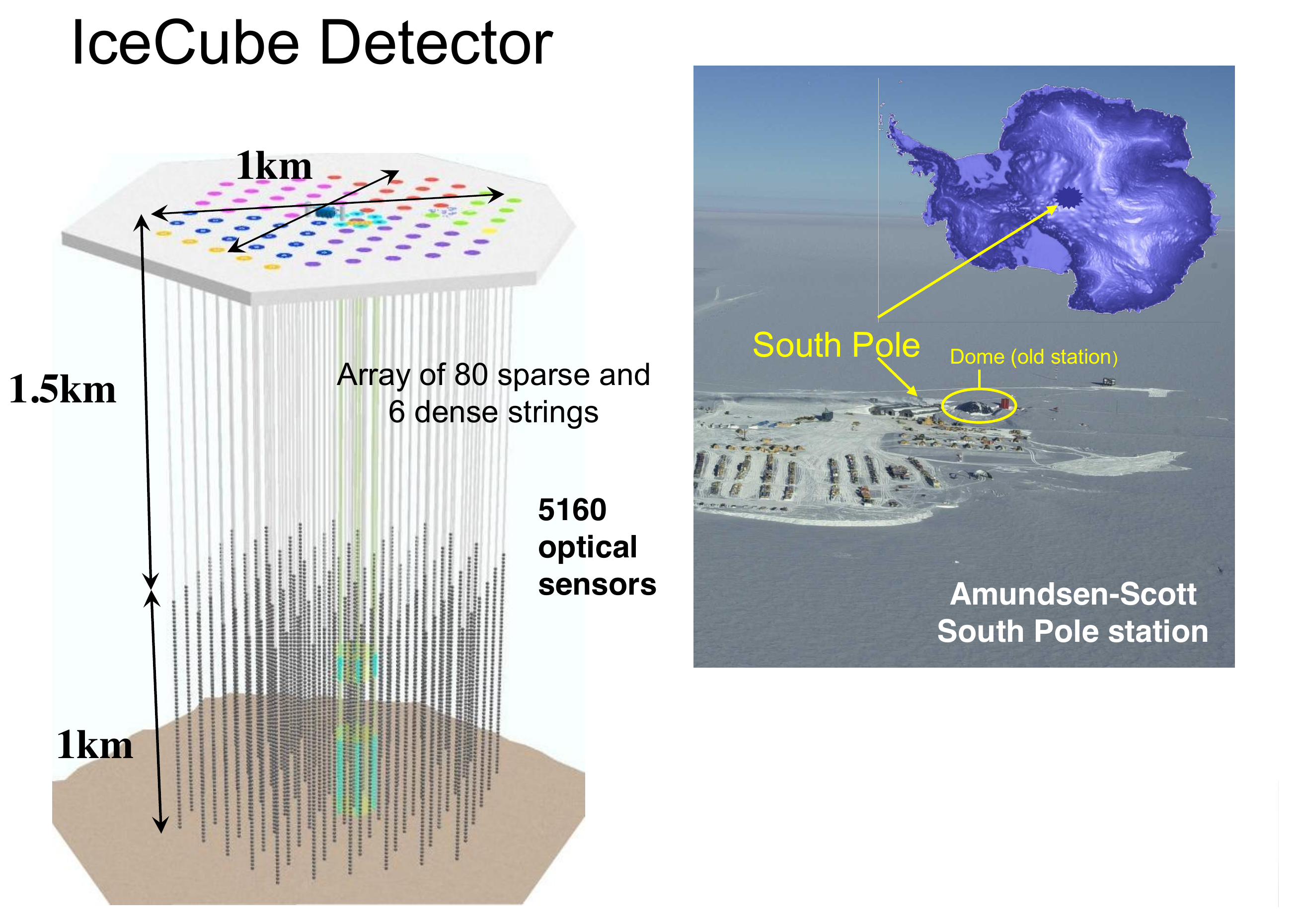}
\end{center}
\caption{Current configuration of the IceCube high-energy neutrino detector.  Figure courtesy of the IceCube Project \cite{Ishihara}.}
\label{fig:icecube}
\end{figure}

A second issue is the existence of point sources that might be probed in neutrinos. 
At very high energies the trajectories
of protons and nuclei are not strongly perturbed by magnetic fields, so that Pierre Auger
results can inform high-energy neutrino studies about potential points sources.
The observations and their interpretation in terms of the GZK effect suggest that the
closest sources of UHECRs are within the GZK volume of $\lsim 100$ Mpc, where the
distribution of matter is inhomogeneous, and thus where there should be anisotropies
in the UHECR flux \cite{PA11}.  Early results from the  Pierre Auger Collaboration \cite{pierre}
showed directional correlations between UHECRs and the positions of cataloged
active galactic nuclei (AGNs) on an angular scale of 3.1$^\mathrm{o}$  at 99\% c.l. The
strength of this correlation has diminished somewhat as the Pierre Auger data set
has been enlarged, but the chance probability that the current data are uncorrelated
with cataloged AGN positions remains below 1\% \cite{PA11}

The challenge in the field of UHE neutrinos is to build telescopes with the necessary
sensitivity to see events, given current estimates of the fluxes (see Fig.~\ref{fig:highEnusnew}).
This requires instrumenting very large volumes.   There have been ongoing efforts
to use large quantities of water and ice as detectors, with experiments completed, operating,
or in development using Antarctic ice (AMANDA, IceCube, ANITA, RICE, ARA, ARIANNA), the oceans
(ANTARES, NEMO, NESTOR, KM3Net), and lakes (Baikal), and with detection methods
including optical and coherent radio detection as well as particle production.

IceCube \cite{Halzen,Ishihara}, a project that extended the dimensions high-energy
neutrino detectors to a cubic
kilometer, has been in operation at the South Pole for the past five years  (Fig.~\ref{fig:icecube}).   This
telescope views the ice through approximately 5160  optical sensors,
deployed on an array of 80 sparse and 6 dense vertical strings, at a depth of 1450 to 2450m.  
The detector is also coupled to a surface air-shower array.  As the earth is opaque to UHE
neutrinos, detection must come from neutrinos incident at or above the horizon.
IceCube deployment began in 2006, and the full set of 86 strings was completed in 2011:
results from 354 days of 86-string operations were reported at Neutrino 2012 \cite{Ishihara}.
To date two high-energy events have been seen, consistent with cosmogenic sources
(e.g., the GZK mechanism or on-site production from a cosmic accelerator), with energies
most likely between 1-10 PeV (1 PeV = $10^{15}$ eV).   It is also possible, though IceCube experimentalist argue
unlikely, that these events are atmospheric
neutrinos. IceCube sensitivity has reached
the upper range of the high-energy neutrino fluxes predicted in cosmogenic neutrino
models.

\section{Acknowledgement}
This work was supported in part by the U.S. Department of Energy
under contracts DE-SC00046548 (UC Berkeley) and
DE-AC02-98CH10886 (LBNL).

\section{Glossary}

\noindent
{\it Atmospheric neutrinos:}\\
Neutrinos produced when cosmic rays strike the Earth's atmosphere.\\

\noindent
{\it Big Bang:}\\
The cosmological model of the origin of the universe in which the current universe resulted from
the expansion of a primordial state characterized by very high temperatures and particle densities.\\

\noindent
{\it Borexino:}\\
A currently operating solar neutrino detector that employs scintillator to detector low-energy
solar neutrinos, such as the ${}^7$Be and pep neutrinos.  Borexino is located in Italy's
Gran Sasso Laboratory.\\

\noindent
{\it CNO cycle:}\\
A series of thermonuclear reactions in which pre-existing C, N, and O act as ``catalysts"  for the conversion of four protons to helium, with release of energy.  The CNO cycle is the dominant
mechanism for ``proton burning" in rapidly evolving, high-mass stars.\\

\noindent
{\it Chlorine Detector:}\\
A solar neutrino detector that operated in the South Dakota Homestake Mine for three decades.
This radiochemical detector contained about 615 tons of chlorine-bearing cleaning fluid.
The discrepancy between the rate of neutrino reactions in this detector and the predictions
of the SSM came to be known as the solar neutrino problem.\\

\noindent
{\it Core-collapse or Type II supernova:}\\
A type of supernova in which a massive star collapses under its own gravity to form a neutron star or black hole, with ejection of the star's mantle, producing a spectacular visual display.  Core-collapse supernovae produce enormous bursts of neutrinos of all flavors.  \\

\noindent
{\it Cosmic microwave background:}\\
The sea of low-energy photons that fills space, left over from the Big Bang.  The spectrum
is that of a black body characterized by a temperature of 2.725K.\\

\noindent
{\it Cosmic rays:}\\
Energetic particles that are produced in various astrophysical sources,
propagate through the interstellar medium, and can interact in
the Earth's atmosphere.  Sources include the winds that blow off stars and the ejecta from
supernovae.  About 90\% of cosmic rays are protons, 9\% are helium or other nuclei, and
1\% are electrons.\\

\noindent
{\it Dark matter:}\\
The term in cosmology given to the largely unidentified matter that influences the large-scale
structure of the universe through its gravitational interactions, but so far has not been detected 
by more direct means.  The majority of matter in our universe is dark.
While neutrinos produced in the Big Bang account for a portion of the dark matter, the bulk
of this matter appears to be something new, not currently described in the Standard Model
of particle physics.\\

\noindent
{\it Dirac neutrino:}\\
A neutrino that has a distinct antiparticle, and thus has a total of four components.
Lepton number is the ``charge" that is used to distinguish the neutrino from the antineutrino.\\

\noindent
{\it IceCube Neutrino Observatory:}\\
A neutrino telescope under construction in Antarctica in which approximately a
cubic kilometer of deep ice will be instrumented as an observatory for high-energy
astrophysical neutrinos.\\

\noindent
{\it Majorana neutrino:}\\
A neutrino that serves as its own antiparticle, and thus has only two components.\\

\noindent
{\it MSW mechanism:}\\
The mechanism by which surrounding matter alters neutrino oscillation probabilities.  In particular, the MSW mechanism can lead to large matter oscillation probabilities even
when the oscillation probabilities in vacuum are small.  The mechanism is
named after the investigators who first discovered the phenomenon (S. Mikheyev and 
A. Smirnov) and first described matter's influence on neutrino mass (L. Wolfenstein).\\

\noindent
{\it Neutrino:}\\
A chargeless elementary particle that interacts only through the weak interaction, carries a spin
of one-half, and has a very small mass.  Neutrinos are produced in a variety of weak interactions, including nuclear beta decay.  Neutrinos come in three types, or flavors: electron, muon, and tauon.\\

\noindent
{\it Neutrino (flavor) oscillations:}\\
The quantum mechanical phenomenon in which a massive neutrino produced in one flavor
state is later found, when detected some distance from the neutrino source, to be
in a different flavor state.\\

\noindent
{\it Neutrino process:}\\
Nucleosynthesis by direct interactions of neutrinos in the mantle or wind of a core-collapse
supernova.\\

\noindent
{\it Neutron star:}\\
The very dense compact object produced from the gravitational collapse of a massive star,
made up primarily of neutrons and supported by the nuclear equation of state (in contrast to
electron-gas interactions that support ordinary stars of
ordinary density).\\

\noindent
{\it Nucleosynthesis:}\\
The production of new nuclei in cosmology and astrophysics.  Nucleosynthesis sites
include the Big Bang, the cores of ordinary stars, the interstellar medium through which
high-energy cosmic rays pass, and explosive astrophysical environments such as
supernovae.\\

\noindent
{\it Pierre Auger Cosmic Ray Observatory:}\\
A new international observatory for the study of ultra-high-energy 
cosmic-ray interactions in the Earth's atmosphere.  The observatory is located in
Argentina's Mendoza Province.\\

\noindent
{\it pp chain:}\\
A set of thermonuclear reactions that, in the cores of stars like our sun, is responsible for the 
conversion of four protons to helium, with release of energy.  The pp chain is the dominant
mechanism for such ``proton burning" in slowly evolving, low-mass stars.\\

\noindent
{\it Red giant:}\\
A later stage of evolution for low- or intermediate-mass stars characterized by large
radii, low surface temperatures, and high luminosities.  One class of red giants
has a helium core supported by electron degeneracy pressure, surrounded by a shell where
protons are burned to helium, adding to the core mass.\\

\noindent
{\it r-process:}\\
The process by which approximately half of the nuclei heavier than iron are synthesized,
including all of the transuranics.  The r- or rapid-neutron-capture process requires enormous
neutron fluences and high temperatures, such as those produced in supernovae,
neutron star mergers, or other explosive astrophysical
environments.\\

\noindent
{\it Solar neutrinos:}\\
Neutrinos produced in the core of the Sun as a byproduct of the thermonuclear reactions of the
pp chain and CNO cycle.\\

\noindent
{\it Sudbury Neutrino Observatory (SNO):}\\
A detector that operated deep underground in a nickel mine located in Sudbury,
Canada.  The inner portion 
of this detector contained one kiloton of heavy water, a target that generates signals
sensitive to neutrino flavor.\\

\noindent
{\it Super-Kamiokande:}\\
The massive (50,000 ton) water detector that operates underground in Japan's Kamioka Mine,
recording solar and atmospheric neutrino interactions. \\

\noindent
{\it Standard solar model (SSM):}\\
The theoretical model of the evolution of the Sun, based on the standard theory of the 
structure of hydrogen-burning stars.  Our detailed knowledge of the Sun $-$ its age, mass,
composition, vibrational modes, etc. $-$ make it an interesting testing ground for
stellar evolution theory.\\

\noindent
{\it Ultra-high-energy cosmic rays (UHECRs):}\\
Cosmic rays with energies of, typically, 10$^{15}$ to 10$^{20}$ eV, a range associated
with very large cosmic accelerators and bounded, on the upper end, by an energy
cutoff associated with cosmic-ray interactions with the cosmic microwave background.\\

\end{document}